\documentclass[12pt]{article}

\usepackage[english]{babel}
\usepackage[utf8]{inputenc}
\usepackage{amsmath}
\usepackage{graphicx}
\usepackage{caption}
\usepackage{subcaption}
\usepackage[colorinlistoftodos]{todonotes}
\usepackage{setspace}
\usepackage[natbibapa]{apacite}
\usepackage[hyphens]{url}
\usepackage{color}

\usepackage[margin=1in]{geometry}

\title{(A) Data in the Life:  Authorship Attribution of Lennon-McCartney Songs}

\author{Mark E. Glickman\thanks{Address for correspondence:
Department of Statistics, Harvard University, 1 Oxford Street,
Cambridge, MA  02138, USA.
E-mail address:  {\tt glickman@fas.harvard.edu}.
Jason Brown is supported by 
NSERC grant RGPIN 170450-2013.
The authors would like to thank Xiao-Li Meng, Michael Jordan,
David C.\ Hoaglin,
and the three anonymous referees for their helpful comments.}
\\
Department of Statistics\\
Harvard University\\
{\tt glickman@fas.harvard.edu}
\and
Jason I. Brown\\
Department of Mathematics and Statistics\\
Dalhousie University\\
{\tt jason.brown@dal.ca}
\and
Ryan B. Song\\
School of Engineering and Applied Sciences\\
Harvard University\\
{\tt ryan.b.song@gmail.com}}

\date{}
\newcommand{\newc}{\newcommand}
\newc{\bc}{\begin{center}}
\newc{\ec}{\end{center}}
\newc{\be}{\begin{enumerate}}
\newc{\ee}{\end{enumerate}}
\newc{\bi}{\begin{itemize}}
\newc{\ei}{\end{itemize}}
\newc{\bd}{\begin{description}}
\newc{\ed}{\end{description}}
\newc{\und}[1]{\underline{#1}}
\newc{\E}{\mbox{E}}
\newc{\V}{\mbox{V}}
\newc{\We}{\mbox{We}}
\newc{\N}{\mbox{N}}
\newc{\B}{\mbox{Bin}}
\newc{\Bern}{\mbox{Bern}}
\newc{\Po}{\mbox{Po}}
\newc{\IG}{\mbox{IG}}
\newc{\Gam}{\mbox{Gam}}
\newc{\bdp}{\mbox{\boldmath $p$}}
\newc{\bdphat}{\hat{\mathsf{p}}}
\newc{\odds}{\mbox{odds}}
\newc{\OR}{\mbox{OR}}
\newc{\stderr}{\mbox{s.e.}}
\newc{\logit}{\mbox{logit}}
\newc{\sign}{\mbox{sign}}
\newc{\SD}{\mbox{SD}}
\newc{\bdmu}{\mbox{\boldmath $\mu$}}
\newc{\bdSigma}{\mbox{\boldmath $\Sigma$}}
\newc{\bdLambda}{\mbox{\boldmath $\Lambda$}}
\newc{\bdmuhat}{\mbox{\boldmath $\hat{\mu}$}}
\newc{\bdeta}{\mbox{\boldmath $\eta$}}
\newc{\bdtheta}{\mbox{\boldmath $\theta$}}
\newc{\bdbeta}{\mbox{\boldmath $\beta$}}
\newc{\bdgamma}{\mbox{\boldmath $\gamma$}}
\newc{\bdetahat}{\mbox{\boldmath $\hat{\eta}$}}
\newc{\bdbetahat}{\mbox{\boldmath $\hat{\beta}$}}
\newc{\bdgammahat}{\mbox{\boldmath $\hat{\gamma}$}}
\newc{\bdthetahat}{\mbox{\boldmath $\hat{\theta}$}}
\newc{\bdvareps}{\mbox{\boldmath $\varepsilon$}}
\newc{\bdzero}{\mbox{\boldmath $0$}}
\newc{\bdone}{\mbox{\boldmath $1$}}
\newc{\bdnu}{\mbox{\boldmath $\nu$}}
\newc{\bdell}{\mbox{\boldmath $\ell$}}
\newc{\bdxi}{\mbox{\boldmath $\xi$}}
\newc{\bdomega}{\mbox{\boldmath $\omega$}}
\newc{\bdepsilon}{\mbox{\boldmath $\varepsilon$}}
\newc{\bdI}{\mathbf{I}}
\newc{\bdP}{\mbox{\boldmath $P$}}
\newc{\bdX}{\mbox{\boldmath $X$}}
\newc{\bdA}{\mbox{\boldmath $A$}}
\newc{\bdB}{\mbox{\boldmath $B$}}
\newc{\bdC}{\mbox{\boldmath $C$}}
\newc{\bdD}{\mbox{\boldmath $D$}}
\newc{\bdG}{\mbox{\boldmath $G$}}
\newc{\bdJ}{\mbox{\boldmath $J$}}
\newc{\Ktil}{\tilde{K}}
\newc{\Khat}{\hat{K}}
\newc{\bda}{\mbox{\boldmath $a$}}
\newc{\bdb}{\mbox{\boldmath $b$}}
\newc{\bdc}{\mbox{\boldmath $c$}}
\newc{\bde}{\mbox{\boldmath $e$}}
\newc{\bdu}{\mbox{\boldmath $u$}}
\newc{\bdv}{\mbox{\boldmath $v$}}
\newc{\bdx}{\mbox{\boldmath $x$}}
\newc{\bdy}{\mbox{\boldmath $y$}}
\newc{\bdz}{\mbox{\boldmath $z$}}
\newc{\bdr}{\mbox{\boldmath $r$}}
\newc{\bdQ}{\mbox{\boldmath $Q$}}
\newc{\bdR}{\mbox{\boldmath $R$}}
\newc{\bdY}{\mbox{\boldmath $Y$}}
\newc{\bdT}{\mbox{\boldmath $T$}}
\newc{\bdW}{\mbox{\boldmath $W$}}
\newc{\bdWtil}{\tilde{\mbox{\boldmath $W$}}}
\newc{\bdH}{\mbox{\boldmath $H$}}
\newc{\bdL}{\mbox{\boldmath $L$}}
\newc{\bdU}{\mbox{\boldmath $U$}}
\newc{\bdV}{\mbox{\boldmath $V$}}
\newc{\Multinom}{\mbox{Multinom}}
\newc{\Var}{\mbox{Var}}
\newc{\var}{\mbox{var}}
\newc{\diag}{\mbox{diag}}
\newc{\thetahat}{\hat{\theta}}
\newc{\tr}{\mbox{tr}}
\newc{\phat}{\hat{p}}
\newc{\Xbar}{\bar{X}}
\newc{\xbar}{\bar{x}}
\newc{\Ybar}{\bar{Y}}
\newc{\ybar}{\bar{y}}
\newc{\dbar}{\bar{d}}
\newc{\yhat}{\hat{y}}
\newc{\bdyhat}{\mbox{\boldmath $\hat{y}$}}
\newc{\ytil}{\tilde{y}}
\newc{\ftil}{\tilde{f}}
\newc{\Ho}{\mbox{\bf H}_o}
\newc{\Ha}{\mbox{\bf H}_a}
\newc{\phatYX}{\phat_Y - \phat_X}
\newc{\SSG}{\mbox{SSG}}
\newc{\SSB}{\mbox{SSB}}
\newc{\SSE}{\mbox{SSE}}
\newc{\SST}{\mbox{SST}}
\newc{\SSR}{\mbox{SSR}}
\newc{\SSAB}{\mbox{SSAB}}
\newc{\MSG}{\mbox{MSG}}
\newc{\MSB}{\mbox{MSB}}
\newc{\MSE}{\mbox{MSE}}
\newc{\MST}{\mbox{MST}}
\newc{\MSAB}{\mbox{MSAB}}
\newc{\dfE}{\mbox{dfE}}
\newc{\dfG}{\mbox{dfG}}
\newc{\dfB}{\mbox{dfB}}
\newc{\dfT}{\mbox{dfT}}
\newc{\dfAB}{\mbox{dfAB}}
\newc{\muhat}{\hat{\mu}}
\newc{\betahat}{\hat{\beta}}
\newc{\alphahat}{\hat{\alpha}}
\newc{\etahat}{\hat{\eta}}
\newc{\phihat}{\hat{\phi}}
\newc{\sigmahat}{\hat{\sigma}}
\newc{\cl}{\centerline}
\newc{\redtitle}[1]{ {\color{red}\und{#1}:} }
\newc{\bluetitle}[1]{ {\color{blue}\und{#1}:} }
\newc{\magentatitle}[1]{ {\color{magenta}\und{#1}:} }
\newc{\R}{\mathbb{R}}
\newc{\trans}{^\mathsf{T}}
\newc{\xtx}{\bdX\trans\bdX}
\newc{\xxtxx}{\bdX(\xtx)^{-1}\bdX\trans}
\newc{\argmin}{\operatornamewithlimits{argmin}}
\newc{\argmax}{\operatornamewithlimits{argmax}}

\begin{document}
\maketitle

\begin{abstract}
The songwriting duo of John Lennon and Paul McCartney, the two founding members 
of the Beatles, composed some of the most popular and memorable 
songs of the last century.  
Despite having authored songs under the joint credit agreement of 
Lennon-McCartney, 
it is well-documented that most of their songs or portions of songs were 
primarily written by exactly one of the two.  
Furthermore, the authorship of some Lennon-McCartney songs is in dispute,
with the recollections of authorship
based on previous interviews with Lennon and
McCartney in conflict.
For Lennon-McCartney songs of known and unknown authorship written and 
recorded over the period 1962-66, 
we extracted musical features from each song or song portion.
These features consist of the occurrence of
melodic notes, chords, 
melodic note pairs, chord change pairs, 
and four-note melody contours.  
We developed a prediction model based on 
variable screening followed by 
logistic regression with elastic net regularization.
Out-of-sample
classification accuracy for songs with known authorship was 76\%, 
with a $c$-statistic from an ROC analysis of 
83.7\%.
We applied our model to the prediction of songs and song portions
with unknown or disputed authorship.
\end{abstract}


\doublespacing

\section{Introduction}

The Beatles are arguably one of the most influential music groups
of all time, having sold over 600 million albums worldwide.
Beyond the initial mania that accompanied their introduction 
to the UK and Europe in 1962-63, and subsequently to the United States in early 1964, 
the Beatles' musical and cultural impact 
still has lasting influence.
The group has been the focus of academic research to an extent that
rivals most classical composers.
\citet{heuger2018} has been maintaining a bibliography 
that contains over~500 entries devoted to academic research
on the Beatles.
Some recent examples of scientific study of Beatles music include
\citet{cathe2016nostalgie} who applied harmonic vectors theory 
to Beatles songs, 
\citet{wagner2003domestication} who analyzed the presence of
blues motifs in Beatles music,
and
\citet{brown2004mathematics} who used Fourier analysis to 
determine the true arrangement and instrumentation of the opening chord 
of ``A Hard Day's Night.''

The songwriting duo of John Lennon and Paul McCartney took the writing
credits for most recorded Beatles songs.  
The two agreed prior to the Beatles' formation that all songs written by 
the two of them, either together or individually, would be credited to
the partnership ``Lennon-McCartney.''
After the Beatles broke up in 1970 and the Lennon-McCartney partnership
dissolved, Lennon and McCartney attempted to clarify their contributions to their
jointly credited songs.
Most often, individual songs were acknowledged to be written entirely by 
either McCartney or Lennon,
though in some cases one would write most of a song and
the other would contribute small portions
or sections of the song.
\citet{compton1988mccartney} 
provided a fairly complete accounting of the 
actual authorships of Lennon-McCartney songs to the extent
they are known through interviews with each of Lennon and McCartney.
According to this listing,
several songs are of disputed musical authorship. 
Some examples include
the entire songs ``Misery,''
``Do You Want to Know a Secret?,'' ``Wait,'' and ``In My Life.''
\citet{womack2007authorship} provided an interesting account
of the discrepancy in Lennon and McCartney's recollection of the 
authorship of ``In My Life'' in particular: 
Lennon wrote the lyrics, McCartney asserted
that he wrote all of the music, and Lennon claimed that McCartney's only 
contribution was helping with the middle eight melody.
Given that further direct questioning about these songs
is unlikely to reveal their true author, 
it is an open question
whether musical features of Lennon-McCartney songs
could provide statistical evidence of song authorship 
between Lennon versus McCartney.

The idea of using statistical models to predict authorship is one that
has been around for over half a century. 
In one of the first successful
attempts at modeling word frequencies,
\citet{mosteller1963inference,mosteller1984applied}
used Bayesian classification models to infer that James Madison 
wrote all of the 12 disputed Federalist papers.
Other recent works related to authorship attribution include 
\citet{efron1976estimating} and \citet{thisted1987did},
who address questions related to Shakespeare's writing, and 
\citet{airoldi2006wrote},
who examine authorship attribution of Ronald Reagan's radio addresses.
Typical text analysis relies on constructing word histograms,
and then modeling authorship as a function of word frequencies.
Basic background on the analysis and modeling of word frequencies
can be found in 
\citet{manning1999foundations}, and these models applied to text
authorship attribution can be found in
\citet{clement2003ngram} and
\citet{malyutov2005authorship}.

This paper is concerned with using harmonic and
melodic information from the corpus of Lennon-McCartney songs
from the first part of the Beatles' career to infer
authorship of songs by John Lennon and Paul McCartney.
It is not unreasonable to assume that Lennon and McCartney 
songs are distinguishable through musical features.
For example, both
\citet{mccormick1998} and \citet{hartzog2016}
observed that Lennon songs have melodies that
tend not to vary substantially in pitch (illustrative examples include
``I Am the Walrus'' and ``Across the Universe''), whereas
McCartney songs tend to have melodies with larger pitch 
changes (e.g., ``Hey Jude'' and ``Oh Darling''). 
However, such anecdotal observations may not sufficiently 
characterize distinctions between Lennon and McCartney --
a more scientific approach is necessary.
Our analyses attempt to capture distinguishing musical features 
through a statistical approach.

Previous work applying quantitative methods to distinguish 
Lennon and McCartney songs is limited.
\citet{whissell1996traditional} performed a stylometric analysis
of Beatles songs based on lyric content via text analyses
to characterize the emotional
differences between Lennon and McCartney over time.
An unpublished paper by
\citet{mcdougal2013} performed a traditional text analysis using
word count methods to compare Lennon and McCartney's lyric usage,
and supplemented the text analysis with auditory-derived information 
from the program The Echo Nest (\url{the.echonest.com}).
More generally, a variety of 
statistical methods for inferring authorship from musical
information have been published.
\citet{cilibrasi2004algorithmic} 
and
\citet{naccache2008learning} 
used Musical Instrument
Digital Interface (MIDI) encoding of songs, which contains information on the
pitch values, intervals, note durations, and instruments to perform 
distance-based clustering.
\citet{dubnov2003using} developed methods to segment 
music using incremental parsing applied to MIDI files in order to 
learn stylistic aspects of music representation.
\citet{conklin2006melodic} also introduced representing
melody as a sequence of segments, and modeled musical style through
this representation.
A different approach was taken by
\citet{george2014computer},
who converted song data into two-dimensional spectrograms, and
used these representations as a means to cluster songs.

Our approach to musical authorship attribution is most closely related to 
methods applied to genome expression studies and other areas
in which the number of predictors is considerably larger than the sample size.
In a musical context, we reduce each song to a vector of binary variables
indicating the occurrences of specified local musical features.
We derive the features based on the entire set of chords that can
be played (harmonic content) and the entire set of notes that can be sung
by the lead singer (melodic content).
From the point of view of melodic sequences of notes or harmonic sequences 
of chords behaving like text in a document,
individual notes and individual chords can be understood as 1-gram
representations.
The occurrence of individual chords and individual notes form an essential 
part of a reduction in a song's musical content.
To increase the richness of the representation, 
we also consider 2-gram representations of chord and melodic
sequences.
That is, we record the occurrence of pairs of consecutive notes and
pairs of consecutive chords as individual binary variables.
Rather than considering larger $n$-gram sequences (with $n>2$)
as a unit of analysis,
we extract local contour information of melodic sequences
indicating the local shape of the melody line
to be a fifth set of variables to represent local features within a song.
Using occurrences of pitches in the sung melodies, 
chords, pitch transitions, harmonic transitions, and contour information
of Lennon-McCartney songs with known authorship
permits modeling of song authorship as a function of musical content.

We 
developed our modeling approach as a two-step algorithm.
First, we kept only musical features that have a sufficiently strong
bivariate association with authorship, an application of
sure independence screening \citep{fan2007variable,fan2008sure}.
With the features that remained, we then
modeled the authorship attribution as a logistic regression,
but estimated the model parameters using elastic net regularization
\citep{glmnet2010,zou2005regularization},
an approach that flexibly constrains the average log-likelihood 
by a convex combination of a ridge penalty 
\citep{le1992ridge}
and a lasso penalty
\citep{tibshirani1996regression,tibshirani2011regression}.
Many other approaches to regularization are possible.
For example, 
\citet{kempfert2018does}, who predict the authorship of Hadyn versus
Mozart string quartets based on musical features,
select their model through subset selection on the 
Bayesian information criterion (BIC) statistic.

This paper proceeds as follows.
We describe the background of the song data collection and formation
in Section~\ref{sec:data}.
This is followed in Section~\ref{sec:model} by the development of
a model for authorship attribution
based on a variable screening procedure followed by
elastic net logistic regression.
The application of the modeling approach is described in
Section~\ref{sec:application} where we summarize the fit of the model
to the corpus of Lennon-McCartney songs of known authorship,
and apply the model results for predicting songs of disputed authorship.
The paper concludes in Section~\ref{sec:discuss} with 
a discussion of the utility of our approach to wider musical settings.
We provide relevant background on musical notes, scales, note intervals,
chords, and song structure
in Appendix~\ref{sec:musicbg}.

\section{Song Data}  \label{sec:data} 

The data used in our analyses consist of melodic and harmonic
information based on Lennon-McCartney songs that were written
between 1962 and 1966.
This period of Beatles music is during the years they toured and
occurred before the band's activities centered on 
studio productions when their songwriting approaches likely changed significantly.
The songs we included in our analyses were
from the original UK-released albums
\textit{Please Please Me,
With the Beatles, A Hard Day's Night, Beatles for Sale, Help!, Rubber
Soul,} and \textit{Revolver}, as well as all the singles from the same era that were not present in any of these albums. 
The essential reference for both the melodic and harmonic content of the songs was \citet{fujita1993beatles},
although the Isophonics online database of chords for 
The Beatles songs 
(\url{http://isophonics.net/content/reference-annotations})
provided additional points of reference for each song.

The authorship of each Lennon-McCartney song, or whether the
authorship credit was in dispute, has been documented in 
\citet{compton1988mccartney},
though for some songs we have found other 
documentation of song authorship.
Aside from recording whether entire songs were written by Lennon
versus McCartney, 
Compton also notes that in many cases songs
had multiple sections with possibly different authors.
For example, the song
``We Can Work It Out'' is credited to McCartney as the author, though 
the bridge section starting with the lyric
``Life is very short...'' was written by Lennon.
In our analyses, we treat these sections as two different
units of analysis with different authors.
Furthermore, several songs that were acknowledged as full collaborations
between Lennon and McCartney were excluded from the corpus of known authorship
from which we develop our prediction algorithm.
The song ``The Word'' is such an example of a full collaboration.
It is plausible that some of the disputed songs were actually
collaborations, but the current information about the songs
did not permit these joint attributions.
The total number of Lennon-McCartney songs or portions of songs 
with an undisputed individual author (Lennon or McCartney) was 70.
Eight songs or portions of songs in this period were of disputed
authorship.

Our process was to manually code 
each song's harmonic (chord) and melodic progressions.
The song content that serves as the input to our modeling strategy
is a set of representations of simple melodic and harmonic patterns 
within each song in the form of category indicators. 
That is, we let each song be represented by a vector of binary variables
within the song,
where each variable is the presence/absence of a musical feature 
that could occur in the song. 
We describe these representations in more detail below.
The process to obtain these category indicators involves converting each
song's melodic and harmonic content into a usable form.
Melody lines were partitioned into phrases which were
typically book-ended by rests (silence).
An alternative approach would have been to model 
counts of musical features within songs, 
which is much more in line with authorship attribution analysis
for text documents.
A crucial difficulty with this approach is 
how to address repeated phrases (e.g., verses, choruses)
within a song.
As an extreme example that is not part of our sample, 
consider
the later-Beatles period McCartney-written song ``Hey Jude.''
The ``na na na'' fadeout, which lasts roughly four minutes on
the recording, is repeated 19 times \citep{everett1999beatles}.
Keeping these repeated occurrences would likely over-represent
the musical ideas suggested by the phrase.
We explored models in which feature counts were incorporated,
including versions where the counts were capped at an upper limit
(i.e., winsorizing the larger counts), and versions involving
the transformation of counts
to the log scale, but these approaches resulted in worse predictability
than our final model.
The use of whether a musical feature was present in a song
produced better discriminatory power in authorship predictions.

The key of every song was standardized relative to the tonic
for songs in a major key, and to the relative major (up a minor third)
for songs in a minor key.
If a key change occurred in the middle of the song, 
the harmonic and melodic information from that point onward would
be standardized to the modulated key.

We constructed five different sets of musical features
within each song as follows
based on processed melodic and harmonic data for the collection 
of songs.
The first set of features was chord types.
Seven diatonic chords, that is, I, ii, iii, IV, V, vi, vii, which
are conventionally the building blocks for most popular Western music,
were their own categories. 
The true diatonic chord on the seventh 
note of the scale is a diminished chord, which was only used once, 
in ``You Won't See Me,'' while the minor vii was 
used more often.
We therefore took the liberty of using the minor vii instead as 
our ``diatonic'' chord on the seventh.
Because diminished and augmented chords were used rarely in general,
we collapsed all occurrences of non-diatonic major chords 
along with augmented chords
into a single category, and
non-diatonic minor chords along with diminished chords
into a single category.
This resulted in a total of 9 categories.
We explored other category divisions, including fewer instances
of collapsed categories, but the sparsity of the data across
the non-diatonic, augmented, and diminished chords resulted in 
less reliable predictability.
Additionally, we decided to group all seventh and extended chords 
(e.g., ninth chord, eleventh chord) with their unaltered 
triad counterparts.

The second set of features consisted of melodic notes.
The octave in which a melodic note was sung was ignored in 
the construction, so that the number of note categories
totaled 12 (the number of pitch classes on the chromatic scale).

The third set of features comprised chord transitions, that is,
pairs of consecutive chords.
As with individual chord categories, considering all combinations
of chord transitions would have resulted in an 
unnecessarily large number of sparsely counted categories.
We collapsed the chord categories as follows.
Each transition among the tonic, sub-dominant (major fourth),
and dominant (major fifth) was
its own category.
Every other transition from a diatonic chord to another diatonic chord,
regardless of the order of the two chords,
was its own category.
For example, transitions from ii to V were grouped with transitions from V to ii.
Transitions that involved the tonic 
and any non-diatonic chord were grouped into one category,
and transitions that involved the dominant 
and any non-diatonic chord were also all grouped into one category.
Chord transitions starting with any non-diatonic chord, and
ending with a diatonic chord (other than the tonic or dominant) was
its own category, and
chord transitions ending with any non-diatonic chord, and
starting with a diatonic chord (other than the tonic or dominant) was
its own category.
Finally, all chord transitions between two non-diatonic chords
fell under one category.
The total number of chord transition categories totaled 24
with these raw category collapsings.
Empty categories from the canon of songs were ignored.

The fourth set of features involved melodic note transitions
as pairs of notes.
In contrast to the single melodic note categories, we considered
the octave of the second note in the pair. 
Thus, each melodic note in a pair could be 
in a three-octave range.
In addition, we considered the start and end rest of a phrase to
be considered a note in constructing note transition categories.
Thus a single note at the start or at the
end of a phrase was each treated as a note transition.
Each start of a phrase on any diatonic note was its own
category, and each end of a phrase on any diatonic note 
was its own category.
All notes on the diatonic scale transitioning from or to the tonic 
was its own category.
Any transition from a pitch on the diatonic or pentatonic scale 
(which includes the flat 3 and flat 7)
to another pitch on the diatonic or pentatonic scale,
including the same pitch, was its own category, regardless of octave.
Upward movements by 2, 3, 4, or 5 notes on the diatonic scale
were individual categories,
and the corresponding downward movements were their own categories.

We performed 
a greater amount of collapsing of categories of melodic transitions
when at least one note in the transition was not on the diatonic scale.
All transitions between the two same non-diatonic notes (excluding the
flat 3 and flat 7) were collapsed into the same category.
All melodic phrases starting on a non-diatonic note were collapsed into
the same category, and all melodic phrases ending on a non-diatonic note
were collapsed into the same category.
A semitone upward or downward movement from a diatonic note to a non-diatonic note
formed two distinct categories, as did a semitone upward or downward movement
from a non-diatonic note to a diatonic note.
All upward movements of at least two semitones involving a non-diatonic note
were collapsed into the same category, and all downward movements of at
least two semitones were collapsed into the same category.
The total number of nonempty categories 
of melodic transitions under this collapsing scheme was~65.
It is worth noting that we had also considered an alternative set 
of melodic transition variables.
These were based to a large extent on grouping upward and downward
movements by the size of the interval, but without regard to the 
musical function of the transition.
We feel that
the main groupings described above are arguably more musically justifiable
because they are more directly connected to the pitches within 
transition pairs rather than pitch distances.

The last set of features captured local contours in the melodic line
of a song.
Every consecutive 4-note subset within a melodic phrase (between its
start and end) was partitioned into one of 27 different categories
according to the direction of each consecutive pair of notes.
For each of the three pairs of consecutive notes in a 4-note melodic
sequence, the transitions could be up, down or same if the melodic
notes moved up, down, or stayed the same.
Because each consecutive pair across the 4-note sequences
allowed three possibilities, the representation consisted
of $3\times 3\times 3 = 27$ categories.
Longer contours (consecutive note subsets of 5 or more notes)
would provide greater contour detail, but the number of implied
categories would create difficulties in model fitting especially with the relatively 
low number of songs to use for model-building.
The contour representation is an attempt to characterize
local features in the melodic line beyond 2-gram representations
but without the same level of detail.

The five sets of musical features
together total 137 binary variables for each song.
Our modeling approach, which relies mainly on cross-validating
regularized logistic regression, can result in 
prediction instability when a feature is shared by very few 
or very many songs. 
We therefore removed 16 features in which five or fewer songs contained
the feature, or where 66 or more songs (out of 70) contained the feature.
The features shared by 66 or more songs included the tonic chord;
melodic notes that included the tonic, second, third and fifth;
and the 4-note contour (up, down, down).
The features shared by five or fewer songs consisted of
the minor seventh chord, 
chord transition from iii to V, upward and downward melodic transitions by 5 notes
on the diatonic scale, repeated flat 3 notes, other repeated non-diatonic notes, 
upward melodic transition from flat 7 to flat 3, 
melodic transition between flat 3 and fifth,
and melodic transition from flat 7 to fourth.
With these exclusions, our analyses used a total of 121 musical features.

We display the most common musical characteristics by 
category, after the exclusions,
in Table~\ref{tbl:common-features}.
Major 4th and major 5th chords are the most common among the 70 songs
(after the tonic), and the melodic notes of a 4th and 6th are
also common.
These notes and chords are understood to be the building blocks
of popular Western music.
The chord transition from major 5th to tonic is also a common
chord change in popular music, 
is well-represented in early Lennon-McCartney songs,
and is often utilized as a harmonic phrase resolution.
The most common melody note transitions stay on the diatonic scale,
which again is in keeping with Western songwriting.
Finally, the two contours listed in Table~\ref{tbl:common-features}
are both simple shapes in the melodic line.
\begin{table}[ht]
\centerline{
\begin{tabular}{l|l}
Representation & Features \\ \hline\hline
Chords & Major 4th (64), Major 5th (63) \\ \hline
Melodic notes & 4th (62), 6th (63) \\ \hline
Chord transitions & Major 5th to Tonic (61) \\ \hline
Note transitions & Downward transition of one note on the diatonic scale (62),\\
 & Upward transition of one note on the diatonic scale (60)\\ \hline
Contours & (down, down, down) (61), (down, down, up) (62) \\ \hline
\end{tabular}
}
\caption{\label{tbl:common-features}
Musical features among the 121 that occurred in 60 or more
of the 70 songs with known authorship, 
after eliminating features occurring in 65 or more songs.
Numbers in parentheses indicate the number of songs with
the listed feature.
}
\end{table}

\section{A model for songwriter attribution} \label{sec:model}

Our approach to modeling authorship involved 
a two-step process.
First, we selected a subset of the 121 musical features that 
each had a sufficiently strong bivariate association with authorship.
Second,
conditional on the selected features,
we modeled authorship using 
logistic regression regularized via elastic net penalization
\citep{zou2005regularization} with tuning parameters
optimized by cross-validation.
The latter process was implemented in the~R
package {\tt glmnet} 
\citep{glmnet2010}.
We describe each step in more detail below.

For song $i$, $i=1,\ldots,n$, where $n$ is the number of songs
with known authorship in the training data, let
\begin{equation}
y_i = \left\{
\begin{array}{rl}
0 & \mbox{if song $i$ was written by John Lennon}\\
1 & \mbox{if song $i$ was written by Paul McCartney.}
\end{array}
\right.
\label{eq:y}
\end{equation}
We let $\bdy = (y_1,\ldots,y_n)'$ denote the vector of binary
authorship indicators.
For $j=1,\ldots,J$, 
where $J$ is the total number of dichotomized musical features,
let for each $i=1,\ldots,n$,
\begin{equation}
x_{ij} = \left\{
\begin{array}{rl}
0 & \mbox{if feature $j$ is not observed in song $i$}\\
1 & \mbox{if feature $j$ is observed in song $i$.}
\end{array}
\right.
\label{eq:x}
\end{equation}
We let $\bdX$ denote the $n \times J$ matrix with elements $x_{ij}$,
and let $\bdX_j$ denote the $j$-th column of $\bdX$.

The first step of our procedure is to determine a subset of the 
index set $\{1,2,\ldots,J\}$ in which $\bdX_j$ is
sufficiently associated with authorship.
This can be accomplished by computing odds ratios of the $j$-th binary
feature with authorship and retaining features with an odds ratio
(or its reciprocal)
above a specified threshold.
Equivalently, the selection can be performed by retaining
features in which tests for significant odds ratios
have $p$-values below a specified level.
This pre-processing of features,
known as sure independence screening (SIS),
has been developed and explored by 
\citet{fan2007variable},
\cite{fan2008sure}, and \citet{fan2010sure}.
SIS is more typically employed in settings with a massive number of 
predictors, but in our setting provides a crude but effective way of
reducing the number of features in our final model.
Our final model evaluations exhibit better out-of-sample accuracy
including SIS as a pre-processing step to modeling than omitting
this step, as we describe in Section~\ref{sec:application}.

To implement SIS in our setting,
we computed a $p$-value of a Pearson chi-squared test 
for each $j=1,\ldots,J$,
for the significance of the odds ratio 
in a $2\times 2$ contingency table constructed from $\bdy$ and $\bdX_j$.
When the elements of any of the contingency tables has low counts,
the odds ratio estimate is unstable.
The reference distribution for such settings is 
poorly approximated by a chi-squared distribution,
so we instead simulated test
statistics 10,000 times from the null distribution according to
\citet{hope1968simplified} to obtain more reliable $p$-values.
This procedure is implemented in the {\tt chisq.test} function in base~R.
The $p$-value for each test was then compared to a pre-specified significance
level to determine inclusion for modeling.
See below for a detailed discussion about the specified significance level.

Suppose as a result of the variable screening we retained $K$ variables,
renumbered $1,\ldots, K$.
The second step of the procedure involves
a logistic regression model of the form
\begin{equation}
p_i = \Pr(y_i = 1 | \bdx_i,\beta_0,\bdbeta) = 
\frac{1}{1 + \exp(-(\beta_0+\bdx_i'\bdbeta))} 
\label{eq:logistic}
\end{equation}
where
$\bdx_i = (x_{i1},\ldots,x_{iK})'$, and with model parameters
$\beta_0$ and $\bdbeta= (\beta_1,\ldots,\beta_K)'$.
Given the possibly large number of musical features compared to the number
of songs in our data set,
we fit our logistic regression model through elastic net regularization.
Letting 
\begin{equation}
\ell(\beta_0,\bdbeta | \bdy,\bdX^*)
=
\sum_{i=1}^n \left(
y_i \log p_i + (1-y_i)\log (1-p_i)
\right)
\label{eq:loglik}
\end{equation}
be the log-likelihood of the model parameters,
where $\bdX^*$ is the $n\times K$
matrix of $x_{ij}$ retained from variable screening, 
elastic net regularization seeks to find estimates of $\beta_0$
and $\bdbeta$, conditional on $\alpha$ and $\lambda$, 
that minimize
\begin{equation}
f_{EN}(\beta_0,\bdbeta | \bdy,\bdX^*,\alpha,\lambda)
=
-\frac{1}{n} 
\ell(\beta_0,\bdbeta | \bdy,\bdX^*)
+
\lambda \left[
(1-\alpha) \frac{\| \bdbeta \|_2^2}{2} +
\alpha \| \bdbeta \|_1
\right]
\label{eq:loglik-penalized}
\end{equation}
where $\| \bdbeta \|_2^2 = \sum_{j=1}^J \beta_j^2$ and
$\| \bdbeta \|_1 = \sum_{j=1}^J |\beta_j|$, and
$\lambda \geq 0$ and $0\leq \alpha \leq 1$ are tuning parameters. 
When $\alpha=0$, regularization is of the form of a ridge ($L_2$)
penalty, and when $\alpha=1$ the logistic regression is fit with
a Lasso ($L_1$) penalty.

Optimization of the elastic net logistic regression parameters
proceeds as follows.
We consider the equally-spaced grid of values for $\alpha$ in
$\{0.0, 0.1, \ldots, 1.0\}$.
For each candidate value of $\alpha$, we consider
100 candidate values of $\lambda$. 
The choice of these candidate values is described in~\citet{glmnet2010}.
For these $11\times 100 = 1100$ candidate pairs $(\alpha,\lambda)$,
we perform 5-fold cross-validation using the 
negative log-likelihoods evaluated at the withheld fold.
Each fold is constructed by sampling songs stratified by author
so that approximately 20\% of Lennon and 20\% of McCartney songs
are contained in each fold.
This approach preserves the balance in authorship within 
fold relative to the overall sample.
We choose the minimizing pair of $\alpha$ and $\lambda$, and then
minimize the target function in~(\ref{eq:loglik-penalized}) 
over the coefficients $\beta_0$ and $\bdbeta$.
\citet{zou2005regularization} argued for considering the selection
of $\lambda$ based on a 1~standard error rule commonly used in 
regularization procedures, but we found in 
our application that choosing the minimum
value resulted in better predictability.

A natural extension to regularized logistic regression is to
include interactions among the predictors.
Among the difficulties of including all interaction terms 
in a regularized regression is that 
the likely higher degree of sparsity among the interactions 
compared with the individual features
makes it difficult
to identify the important interactions.
Futhermore,
high correlations among
the variables can negatively impact selection.
Work aimed at discovering important interactions
in a more principled manner has been explored.
\citet{ruczinski2003logic,ruczinski2004exploring}
developed logic regression, a procedure that finds Boolean
combinations of binary predictors in an approach similar
to Bayesian CART
\citep{chipman1998bayesian}.
Logic regression prevents overfitting through the reduction of
model complexity in growing the number of Boolean combinations
that are formed.
Procedures such as those by~\citet{bien2013lasso} and~\citet{lim2015learning}
involve building interactions only when the main effect terms are
selected, and this is carried out by taking advantage of the
group-lasso \citep{yuan2006model}.
We explored these extensions to our approach, 
based on having already eliminated the rarely-occurring or 
frequently-occurring musical features,
but found
that out-of-sample
predictability was worse than using only the additive effects of our features.
An argument could be made that including interactions would better account
for sets of musical features that are highly correlated.
However, the extra flexibility associated with including interactions results
in greater variance in the predictions that degrades our model's performance.

Rather than specifying a single
significance level threshold for variable screening followed by 
regularized logistic regression,
our selection procedure considered five different significance level 
thresholds:
1.0 (no variable screening), 0.75, 0.50, 0.25, and 0.10.
We discuss in Section~\ref{sec:discuss} the rationale for only
four additional thresholds.
We performed leave-one-out cross-validation 
in the following manner to choose the
best threshold.
Let $\bdX_{(i)}$ and $\bdy_{(i)}$ denote the predictor matrix and response
vector with observation $i$ deleted.
First, for a fixed threshold $t\in \{1.0, 0.75, 0.50, 0.25, 0.10\}$, 
we performed variable screening on $\bdX_{(i)}$
followed by fitting elastic net logistic regression of $\bdy_{(i)}$
based on the retained features
(with 5-fold cross-validation within the $n-1$ songs to obtain
the elastic net parameter estimates).
The out-of-sample predicted probability $\phat_i^{(t)}$ for observation $i$ 
and threshold $t$
is then computed given $\bdx_i$ from the fitted logistic regression.
The negative log-likelihood for threshold $t$ is computed as
\begin{equation}\label{eq:LL}
\mbox{LL}^{(t)} = -\sum_{i=1}^n \left( y_i \log \phat_i^{(t)}
+ (1-y_i)\log(1-\phat_i^{(t)}) \right) .
\end{equation}
The threshold $t=t_{opt}$ with the minimum $\mbox{LL}^{(t)}$ is the one
chosen by this procedure.
Once $t_{opt}$ is determined, variable screening is performed using this
threshold based on all $n$ observations followed by performing regularized
logistic regression on the remaining features.

\section{Model implementation and results} \label{sec:application}

We applied our approach to authorship attribution 
developed in Section~\ref{sec:model}
to the corpus of 70 Lennon-McCartney songs based on the musical
features described in Section~\ref{sec:data}.
We first describe model summaries applied to the 70 Lennon-McCartney
songs in the training data.
These summaries are based on a leave-one-out predictive analysis.
We then fit our model to the full 70 songs, and use the
results to make predictions on the songs and song portions that are
of disputed authorship or are known to be collaborative.

\subsection{Predictive validity and leave-one-out model summaries} 
\label{subsec:predval}

A common approach to predictive validity in machine learning
is to divide a data set into
modeling, validation, and calibration subsets.
Typically a model is constructed and validated iteratively on the first two
subsets of the sample, 
and predictive properties of the approach
are summarized on the withheld calibration set.
See \citet{draper2013bayesian} for a good overview of this approach,
which the author terms ``calibration cross-validation.''
Given the small number of observations (songs) in our sample,
our predictive accuracy would suffer by withholding a substantial calibration
set, so instead we summarized our algorithm's quality of calibration through
leave-one-out cross-validation.
Specifically, we withheld one song at a time, and with the remaining 69 songs
we performed the procedure described in Section~\ref{sec:model}.
That is, with 69 songs at a time, we first optimized the choice of the
$p$-value threshold for SIS through leave-one-out cross-validation (with a 
68-versus-1 split to compute the out-of-sample negative log-likelihood),
then with the variables selected based on the
optimized $p$-value threshold we fit a logistic regression 
via elastic net regularization
on the 69 songs (using 5-fold cross-validation to estimate the tuning parameters).
Finally, based on the logistic regression fit, the probability estimate
of the withheld song was computed.
This process was performed for all 70 songs to obtain out-of-sample
predictions for each song with known authorship.

\begin{figure}[ht]
\centering
\includegraphics[width=0.9\textwidth]{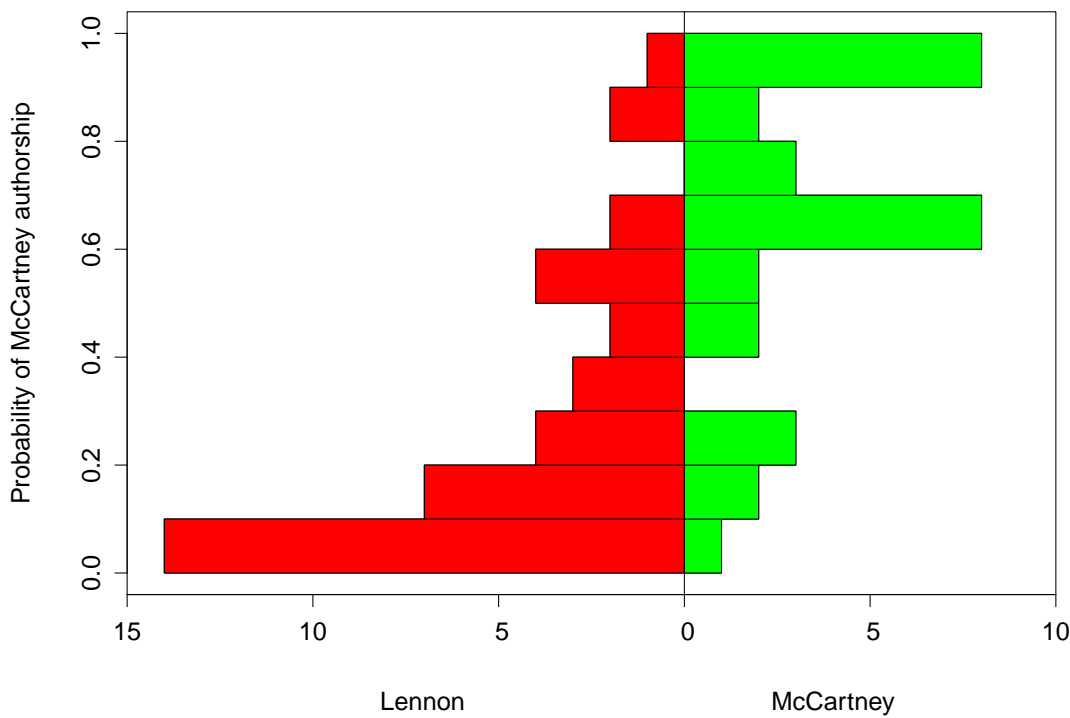}
\caption{\label{fig:hist-back2back}
Back-to-back histograms of the out-of-sample prediction probabilities
of songs of known authorship.
Bars to the left represent 39 songs or song portions
known to be written by Lennon,
and bars to the right represent 31 songs or song portions
known to be written by McCartney.
}
\end{figure}
Figure~\ref{fig:hist-back2back} displays histograms of
the out-of-sample probabilities McCartney wrote each of the~70 songs
or song portions
with known authorship.
The songs or fragments were divided into the 39 that Lennon wrote, and the 31 that
McCartney wrote.
Generally, the higher probability estimates tend to correspond
to McCartney-authored songs, and lower probabilities correspond to
Lennon songs.
Using 0.5 as a threshold for classification, the model
correctly classifies 
76.9\%
of Lennon songs, and 
74.2\%
of McCartney songs,
with an overall correct classification rate of 75.7\%.
We display the 
leave-one-out probability predictions
for the 39 songs known to be written by Lennon in Table~\ref{tbl:johnsongs},
and for the 31 songs known to be written by McCartney
in Table~\ref{tbl:paulsongs}.

\begin{table}[ht]
\centerline{
\begin{tabular}{l|c}
     & McCartney \\
Lennon-authored Song & Probability \\ \hline\hline
All I've Got To Do & 0.008\\ \hline 
Doctor Robert & 0.012\\ \hline 
I'm Happy Just to Dance With You & 0.038\\ \hline 
No Reply & 0.041\\ \hline 
Girl & 0.047\\ \hline 
I'll Be Back & 0.048\\ \hline 
I'm Only Sleeping & 0.049\\ \hline 
There's a Place & 0.064\\ \hline 
I'll Cry Instead & 0.065\\ \hline 
When I Get Home & 0.066\\ \hline 
And Your Bird Can Sing & 0.067\\ \hline 
Help! & 0.071\\ \hline 
We Can Work It Out (Bridge) & 0.071\\ \hline 
You're Going to Lose that Girl & 0.076\\ \hline 
I'm a Loser & 0.100\\ \hline 
Run For Your Life & 0.109\\ \hline 
It's Only Love & 0.111\\ \hline 
This Boy & 0.128\\ \hline 
I Call Your Name & 0.148\\ \hline 
It Won't Be Long & 0.178\\ \hline 
Please Please Me & 0.185\\ \hline 
You Can't Do That & 0.231\\ \hline 
Ticket to Ride & 0.244\\ \hline 
A Hard Day's Night (Verse/Chorus) & 0.279\\ \hline 
Day Tripper & 0.294\\ \hline 
I Don't Want to Spoil the Party & 0.332\\ \hline 
Tomorrow Never Knows & 0.378\\ \hline 
Not a Second Time & 0.390\\ \hline 
Tell Me Why & 0.438\\ \hline 
Nowhere Man & 0.445\\ \hline 
You've Got to Hide Your Love Away & 0.524\\ \hline 
If I Fell & 0.574\\ \hline 
Any Time At All & 0.588\\ \hline 
I Feel Fine & 0.598\\ \hline 
I Should Have Known Better & 0.615\\ \hline 
Norwegian Wood (Verse/Chorus) & 0.666\\ \hline 
Yes It Is & 0.802\\ \hline 
She Said She Said & 0.836\\ \hline 
What Goes On (Verse/Chorus) & 0.944\\ \hline 
\end{tabular}
}
\caption{ \label{tbl:johnsongs}
Songs or song fragments known to be written by
John Lennon, rank ordered according to
the out-of-sample probability (second column) that
is attributed to Paul McCartney.}
\end{table}

\begin{table}[ht]
\centerline{
\begin{tabular}{l|c}
     & McCartney \\
McCartney-authored Song & Probability \\ \hline\hline
You Won't See Me & 0.069\\ \hline 
And I Love Her (Verse/Chorus) & 0.105\\ \hline 
For No One & 0.184\\ \hline 
Here There and Everywhere & 0.202\\ \hline 
PS I Love You & 0.282\\ \hline 
I'll Follow the Sun & 0.284\\ \hline 
Can't Buy Me Love & 0.440\\ \hline 
Got to Get You Into My Life & 0.448\\ \hline 
Eight Days a Week & 0.528\\ \hline 
Eleanor Rigby & 0.570\\ \hline 
I'm Down & 0.606\\ \hline 
Hold Me Tight & 0.606\\ \hline 
She's a Woman & 0.660\\ \hline 
I've Just Seen a Face & 0.668\\ \hline 
Tell Me What You See & 0.668\\ \hline 
What You're Doing & 0.679\\ \hline 
Drive My Car & 0.688\\ \hline 
Yesterday & 0.689\\ \hline 
The Night Before & 0.715\\ \hline 
All My Loving & 0.719\\ \hline 
Yellow Submarine & 0.734\\ \hline 
Every Little Thing & 0.806\\ \hline 
We Can Work It Out (Verse/Chorus) & 0.866\\ \hline 
Michelle (Verse/Chorus) & 0.912\\ \hline 
Things We Said Today & 0.938\\ \hline 
Good Day Sunshine & 0.953\\ \hline 
I'm Looking Through You & 0.957\\ \hline 
Another Girl & 0.964\\ \hline 
I Saw Her Standing There & 0.979\\ \hline 
I Wanna Be Your Man & 0.986\\ \hline 
Love Me Do & 0.989\\ \hline 
\end{tabular}
}
\caption{ \label{tbl:paulsongs}
Songs or song fragments known to be written by
Paul McCartney, rank ordered according to
the out-of-sample probability (second column) that
is attributed to Paul McCartney.}
\end{table}

\begin{figure}[ht]
\centering
\includegraphics[width=0.8\textwidth]{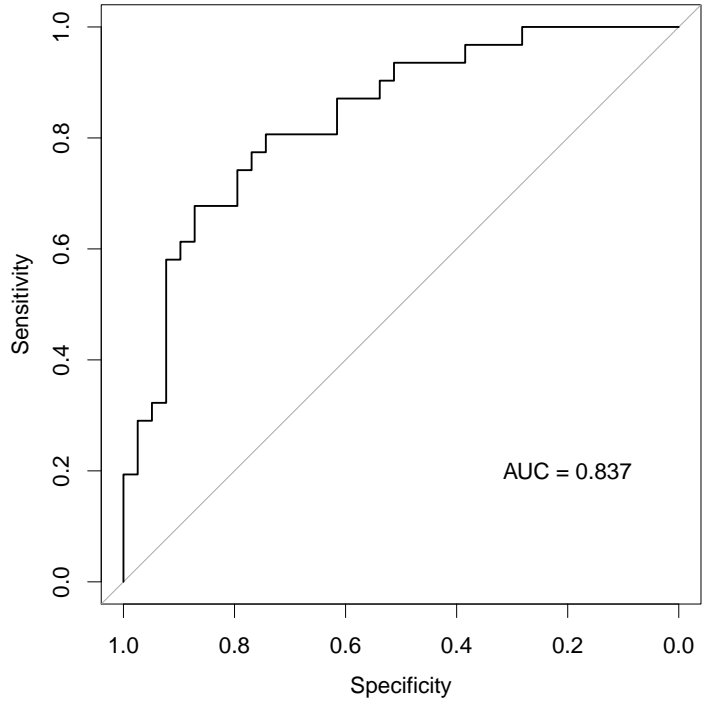}
\caption{\label{fig:rocplot}
ROC plot for out-of-sample song probability predictions
based on 70 songs or song fragments with known authorship.
}
\end{figure}
In addition to the simple classification results, we performed
a receiver operating characteristic curve (ROC) analysis on 
the out-of-sample 
probability predictions for the 70 songs and fragments.
The results of the analysis, which were performed using the
\verb+pROC+ library in R \citep{pROC2011},
are summarized in Figure~\ref{fig:rocplot}.
The $c$-statistic (or area under the ROC curve, AUC)
is 0.837,
which indicates a strong level of 
predictive discrimination.

\begin{figure}[ht]
\centering
\includegraphics[width=0.8\textwidth]{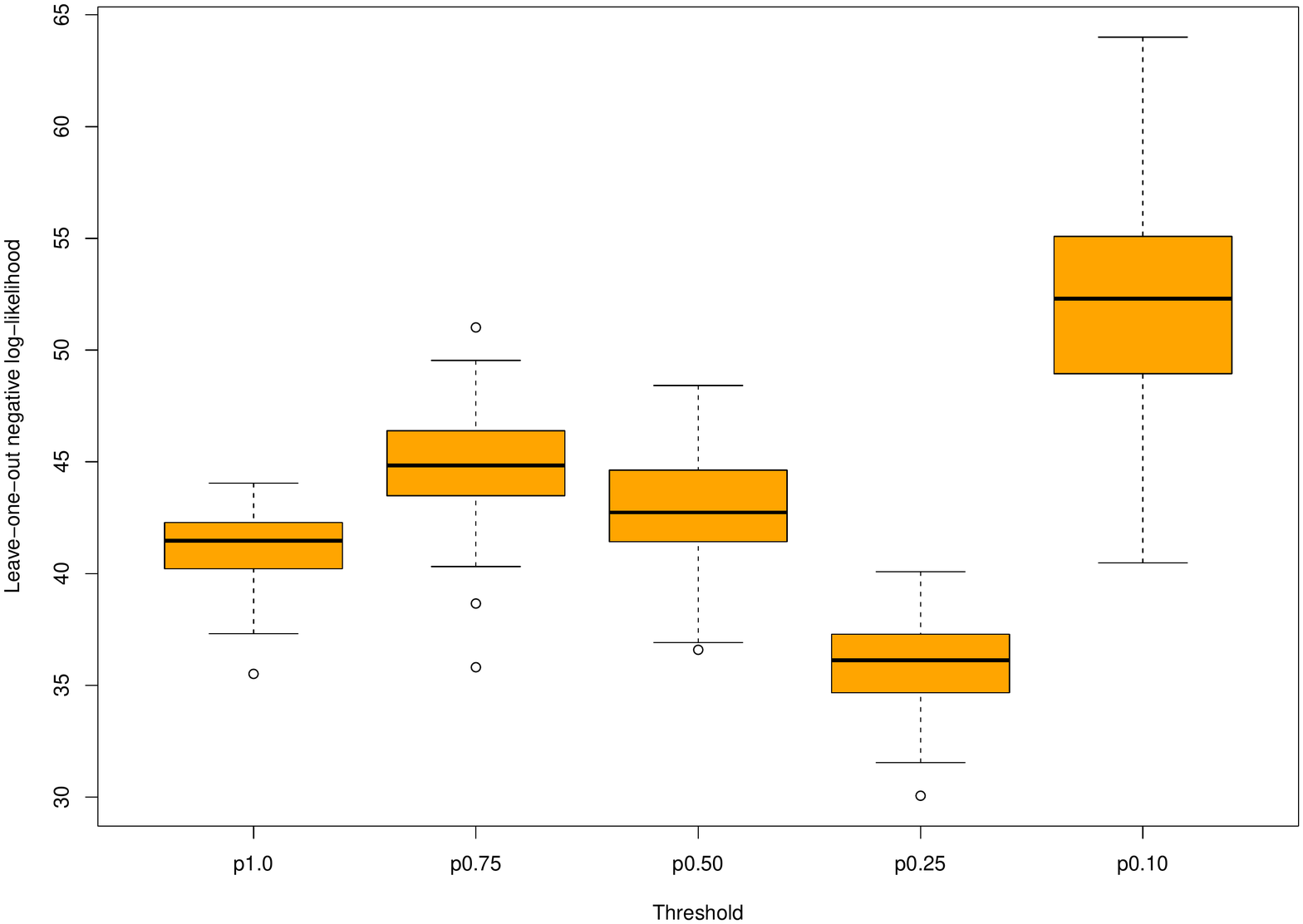}
\caption{\label{fig:boxp}
Boxplots of the leave-one-out negative log-likelihoods
for each choice of $p$-value threshold (1.0, 0.75, 0.5, 0.25, 0.10).
Each box consists of the distribution of negative log-likelihoods
across the 70 leave-one-out analyses.
}
\end{figure}
For each of the 70 applications of optimized variable screening followed by
regularized logistic regression based on 69 songs at a time,
we recorded the optimal variable screening $p$-value threshold.
We discovered that among the $p$-value thresholds in the candidate set,
the significance level of 0.25 was selected in 
69 of the 70 applications of variable screening, and the significance level of 1.0
was selected for one application (corresponding to leaving out the song
``You Won't See Me'' by McCartney).
Figure~\ref{fig:boxp} shows boxplots across the 70 analyses
of the leave-one-out
predictive negative log-likelihoods for each $p$-value threshold.
As seen in the figure, the negative log-likelihoods 
achieve their lowest values when 
discarding features that have a $p$-value for an odds ratio
larger than 0.25.
The second-best choice among these candidate thresholds
was not to remove any variable prior to elastic net.
Removing variables based on a threshold of 0.10 resulted in
noticeably worse performance than any of the other choices.

\subsection{Probability predictions for disputed and collaborative songs}
\label{subsec:dispute}

We applied our algorithm from Section~\ref{sec:model}
to the full set of 70 songs.
The resulting logistic regression model was
then used to make predictions on disputed songs
and song portions, and on songs known to be collaborations
between Lennon and McCartney.
The optimal significance level threshold for the variable screening
was 0.25 based on leave-one-out cross-validation.
Conditional on selecting variables using the 0.25 $p$-value threshold, 
the tuning parameters in elastic net logistic regression 
were optimized at 
$\alpha=0.3$ and $\lambda=0.0359$.
Thus, the final logistic regression model for predictions
involved 
an average of $L_1$ and $L_2$ penalties, but more heavily 
weighted towards a ridge penalty.
Of the 40 features that were selected through sure independence screening,
29 were non-zero in the final model
as a result of elastic net logistic regression.
The full set of 29 variables 
is listed in 
Table~\ref{tbl:allcoefs}. 
\begin{table}[ht]
\centerline{
\begin{tabular}{l|r|r}
Feature & Coefficient & $c$-statistic \\ \hline\hline
Intercept & --0.796 & ---\\ \hline
Chord: V & 1.096 & 0.806\\ \hline
Chord: iii & --0.350 & 0.842\\ \hline
Note:  Flat 2 & --0.874 & 0.817\\ \hline
Note:  Flat 3 & 0.603 & 0.828\\ \hline
Note:  4th & 1.347 & 0.788\\ \hline
Note:  6th & 0.046 & 0.825\\ \hline
Chord transition: between I and vi & --0.315 & 0.823\\ \hline
Chord transition: between ii and iii & --0.255 & 0.846\\ \hline
Chord transition: between ii and IV & 1.428 & 0.795 \\ \hline
Chord transition: between ii and V & --0.291 & 0.830 \\ \hline
Chord transition: non-diatonic to diatonic & --0.096 & 0.833\\ \hline
Melodic transition: down from 4th to flat 3rd & 0.481 & 0.849\\ \hline
Melodic transition: down from flat 3rd to tonic & 1.206 & 0.778\\ \hline
Melodic transition: down 1 note on diatonic scale, not incl. 1 or
$4\rightarrow5/5\rightarrow 4$ & --0.348 & 0.824\\ \hline
Melodic transition: down 1 half step from non-diatonic to diatonic & 1.030 & 0.797\\ \hline
Melodic transition: phrase end on 5th & --0.633 & 0.808\\ \hline
Melodic transition: pair of notes on the 6th & --0.218 & 0.825\\ \hline
Melodic transition: up 1 note on diatonic scale, not incl. 1 or
$4\rightarrow 5/5\rightarrow 4$ & --0.576 & 0.821\\ \hline
Melodic transition: up 1 half step from non-diatonic to diatonic & --1.232 & 0.798\\ \hline
Melodic transition: up from tonic to flat 3rd & 0.376 & 0.833\\ \hline
Melodic transition: from 3rd to tonic & 0.284 & 0.829\\ \hline
Melodic transition: from 4th to 5th & --0.653 & 0.816\\ \hline
Melodic transition: up from or to a non-diatonic note & 1.135 & 0.806\\ \hline
Contour: (Up, Up, Down) & --0.098 & 0.841\\ \hline
Contour: (Down, Down, Same) & 0.535 & 0.824\\ \hline
Contour: (Up, Same, Same)  & --0.098 & 0.835\\ \hline
Contour: (Down, Up, Same) & --0.938 & 0.825\\ \hline
Contour: (Same, Down, Up) & --0.501 & 0.812\\ \hline
Contour: (Up, Down, Up) & --0.555 & 0.826\\ \hline
\end{tabular}
}
\caption{ \label{tbl:allcoefs}
Coefficient estimates in the final logistic regression in the
second column, and ROC analysis
$c$-statistics in the third column.
The $c$-statistics are computed from the 
70 leave-one-out probabilities with the variable removed from
the prediction algorithm;
thus smaller $c$-statistics indicate greater variable importance.
}
\end{table}

Distinguishing song features of Lennon and McCartney authorship
can be learned from the coefficient estimates of the logistic regression.
Positive coefficients are indicative of features used more
associated with McCartney's songs, 
and negative coefficients are indicative of
features more associated with Lennon's songs.

These results offer interesting interpretations of
musical features that distinguish McCartney and Lennon songs.
One clear theme that emerges is that McCartney tended to use
more non-standard musical motifs than Lennon.
For example, 
the harmonic transitions
between I $\rightarrow$ vi and vi $\rightarrow$ I 
are moves that
are natural and reasonably direct in popular music, and Lennon
used these chord changes much more frequently than McCartney
(coefficient of $-0.315$).
These chord changes also create an ambiguity about
whether the song is in the major or relative minor key. 
Lennon songs like ``It's Only Love'' start with two sets
of alternations between I and vi.
In contrast, the chord change between ii and IV
(coefficient of $1.428$)
is less standard,
and is used more frequently by McCartney, as offering a different ``flavor'' 
to the often used sub-dominant, 
and is used, for example, in McCartney's ``I'm Looking Through You.''

Another example is that Lennon's melodic note changes tended to
remain much more often on the notes of the diatonic scale,
whereas McCartney tended to use melodic note transitions that
moved off the diatonic scale.
This is exhibited in the negative coefficients for
note transitions moving up or down one note on the diatonic
scale, and the positive coefficient 
($1.135$)
for upward note transitions
in which one was not on the diatonic scale.
Lennon also more often started melodic phrases at the 3rd
or ended phrases at a 5th, both of which are notes on the
diatonic scale.
In contrast, McCartney more often used a flat 3,
and transitions from the flat 3 to the tonic 
in his sung melodies, both of which are notes often associated
with a blues scale and not the diatonic scale.
This observation is at odds with the often-held notion that Lennon
composed songs in a more traditional ``rock-and-roll'' style.
In general, these results suggest that the greater complexity
in McCartney's music is a distinguishing feature exhibited by
the coefficients in Table~\ref{tbl:allcoefs} that are positive.

In addition to the coefficients, we report a measure of variable
importance in the third column of Table~\ref{tbl:allcoefs}.
Our measure has close connections to an early approach
developed in the context
of random forests \citep{breiman2001random}.
In particular, the importance of a variable can be assessed 
by randomly permuting its values across observations, and then
computing an overall measure of model performance.
The lower the performance measure after permuting the variable, 
the more important the variable.
For our approach, randomly permuting the values of a musical feature
across songs
is effectively equivalent to having the feature removed 
because sure independence scanning should eliminate the feature
in the first step of our prediction algorithm.
Thus, our variable importance measure was computed as follows.
First, we removed the musical feature whose importance we wanted to assess.
We then applied our out-of-sample
procedure from Section~\ref{subsec:predval} 
and computed 70 leave-one-out predicted probabilities.
We performed an ROC analysis on these probabilities and the known
authorship of the 70 songs and summarized the $c$-statistics in the
third column of Table~\ref{tbl:allcoefs}.
Lower values of the $c$-statistic indicate greater variable importance.
The $c$-statistic without eliminating any features is 0.837, but
some of the values in Table~\ref{tbl:allcoefs} can be higher given
the random assignments in the stratified cross-validation procedure.
Generally, higher absolute values of coefficient estimates correspond to lower
$c$-statistics.
Musical features with the lowest $c$-statistics, all less than 0.80, 
include the McCartney features 
(1) the occurrence of the 4th note on the diatonic scale,
(2) the chord transition between ii and IV, 
(3) the note transition downward from the flat 3rd to the tonic, and
(4) the note transition downward a half step from a non-diatonic note
to a diatonic note.
The only feature with a Lennon leaning and having a $c$-statistic less
than 0.80 is the note transition up a half step from a non-diatonic note
to a diatonic note.
Compared with the McCartney feature of a downward half-step move,
upward half-step moves may correspond to particular note transitions
that are distinct from the downward moves.

\begin{figure}[ht]
\centering
\includegraphics[width=0.9\textwidth]{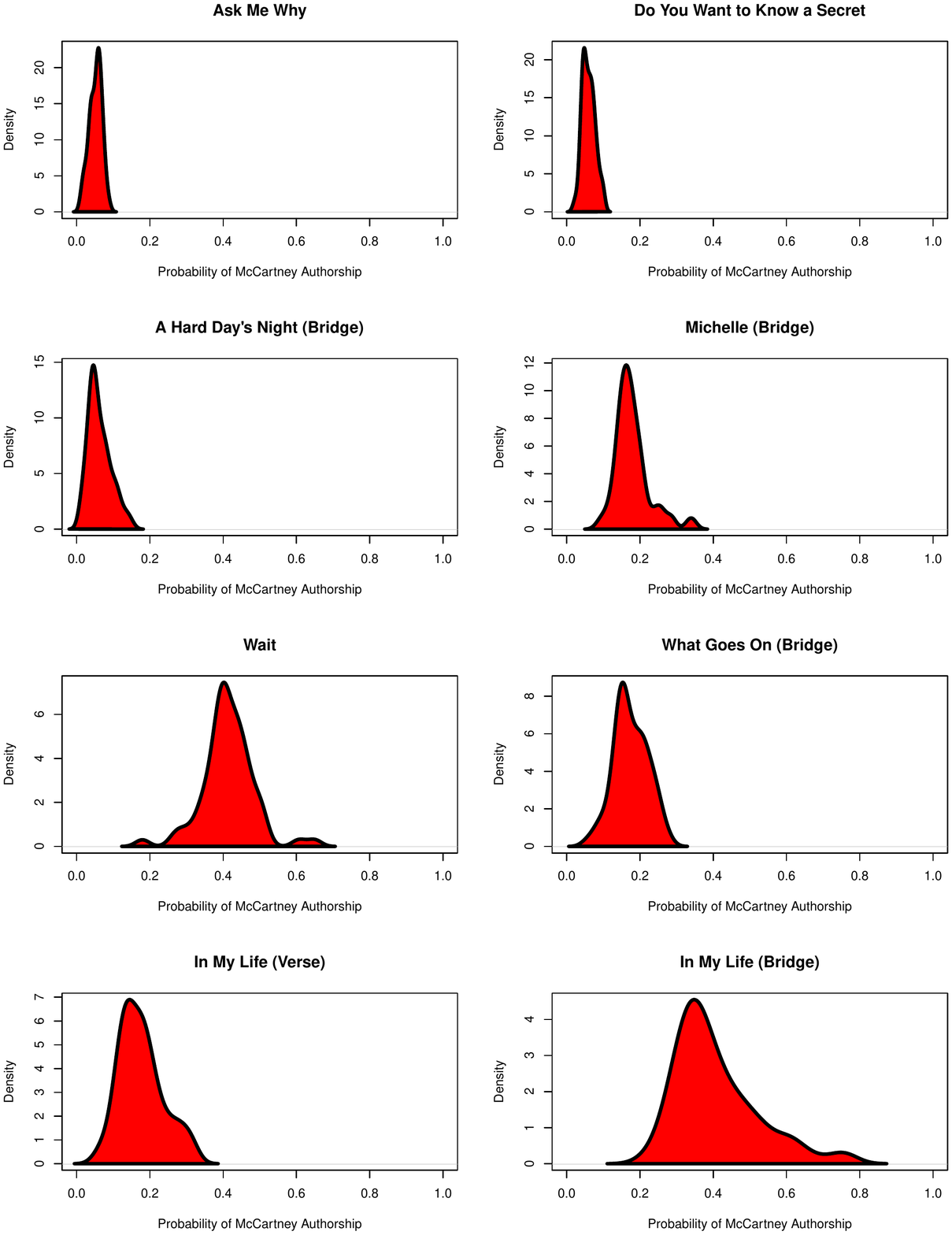}
\caption{\label{fig:density-disputed}
Density plots of the 
leave-one-out probability predictions
for the eight songs of disputed authorship.}
\end{figure}

We applied the fit of our model to make predictions 
for eight songs or song portions with disputed authorship,
and for 11 known to be collaborations.
The prediction probabilities were derived 
by applying the fitted logistic regression to
the songs of unknown and collaborative authorship.
We accompanied the probability predictions with approximate
95\% confidence intervals calculated in the following manner.
For each song of disputed or collaborative authorship, we 
computed 70 probability predictions based on leaving out each one of
the 70 songs in our training sample.
An approximate 95\% confidence interval is constructed from the 
2.5\%-ile and 97.5\%-ile of the 70 probability predictions for each
song.
It is worth noting that these intervals are conservative because 
one fewer song is used than with the corresponding point prediction.
The probability predictions 
and corresponding confidence intervals
are displayed
in Tables~\ref{tbl:songpreds-unknown}
and~\ref{tbl:songpreds-collab}.
\begin{table}[ht]
\centerline{
\begin{tabular}{l|r}
     & McCartney Probability\\
Song & (95\% confidence interval)\\ \hline\hline
Ask Me Why & 0.057 (0.018, 0.080) \\ \hline
Do You Want to Know a Secret & 0.080 (0.033, 0.097) \\ \hline
A Hard Day's Night (Bridge) & 0.069 (0.016, 0.135) \\ \hline
Michelle (Bridge) & 0.199 (0.109, 0.300) \\ \hline
Wait & 0.391 (0.275, 0.540) \\ \hline
What Goes On (Bridge) & 0.235 (0.088, 0.255) \\ \hline
In My Life (Verse) & 0.189 (0.079, 0.307) \\ \hline
In My Life (Bridge) & 0.435 (0.270, 0.692) \\ \hline
\end{tabular}
}
\caption{\label{tbl:songpreds-unknown}
Probability estimates for eight songs or song fragments of disputed or
unknown authorship 
with 95\% confidence intervals
based on a leave-one-out analysis
being attributable to McCartney.
}
\end{table}
We also display the distributions over the 70 predicted probabilities
for each disputed song as density estimates in Figure~\ref{fig:density-disputed}.
For the songs and fragments of disputed authorship, 
all of the probabilities are lower than 0.5 suggesting
that each individually had a higher probability of being
written by Lennon.
The 95\% confidence intervals are mostly less than 0.5,
though ``Wait'' and the bridge of ``In My Life'' have confidence
intervals that cross 0.5.
The density plots in Figure~\ref{fig:density-disputed} demonstrate
the substantial uncertainty in the probability prediction
for the bridge of ``In My Life'' and to a lesser extent for ``Wait.''
In most instances,
the conclusions based on our model seem to match up with 
the suspected authorship, as discussed by \citet{compton1988mccartney}.
According to Compton,
the song ``Ask Me Why,'' which Lennon sang, was likely 
written by Lennon.
Similarly, ``Do You Want to Know a Secret?''\  was one that Lennon recalled
having written and then given to George Harrison to sing.
In ``A Hard Day's Night,'' the verse and chorus are known to
have been written by Lennon \citep{rybac2018,wiener1986beatles},
but McCartney seemed to remember having collaborated, perhaps with the bridge, 
which he sang.
While McCartney wrote most and possibly all of ``Michelle,''
Lennon claimed in some interviews that he came up with the bridge 
on his own, but in other interviews
asserted that the bridge was a collaboration with McCartney
\citep{compton1988mccartney}.
``Wait'' is also suspected to have been written by Lennon
according to \citet{compton1988mccartney}, though in
\citet{miles1998paul} McCartney remembers the song as mostly his.
Lennon wrote ``What Goes On'' several years prior
to the formation of the Beatles, and it is disputed
whether McCartney (and Ringo Starr) helped write the bridge
section when recording the song with the Beatles.
We discuss ``In My Life'' in more detail below.

For the songs during the study period that were
understood to be collaborative, it is unclear
to what extent Lennon and McCartney shared songwriting efforts.
Our model's probability predictions can be viewed 
as demonstrating similarities with the patterns inferred
in songs and fragments with undisputed authorship.
However, it is worth noting that our model was
developed on a set of songs and song portions that
were of single authorship, and that applying our model
to songs of collaborative authorship may 
result in predictions that are not 
as trustworthy.
\begin{table}[ht]
\centerline{
\begin{tabular}{l|r}
     & McCartney Probability\\
Song & (95\% confidence interval)\\ \hline\hline
Misery & 0.310 (0.245, 0.451) \\ \hline
And I Love Her (Bridge) & 0.263 (0.110, 0.315) \\ \hline
Norwegian Wood (Bridge) & 0.330 (0.135, 0.408) \\ \hline
Little Child & 0.337 (0.175, 0.417) \\ \hline
Baby's in Black & 0.920 (0.822, 0.977) \\ \hline
The Word & 0.976 (0.899, 0.994) \\ \hline
From Me To You & 0.606 (0.510, 0.721) \\ \hline
Thank You Girl & 0.106 (0.036, 0.202) \\ \hline
She Loves You & 0.616 (0.515, 0.733) \\ \hline
I'll Get You & 0.062 (0.016, 0.107) \\ \hline
I Want to Hold Your Hand & 0.115 (0.065, 0.182) \\ \hline
\end{tabular}
}
\caption{\label{tbl:songpreds-collab}
Probability estimates for 11 collaborative songs or song fragments 
with 95\% confidence intervals
based on a leave-one-out analysis
being attributable to McCartney.
}
\end{table}
As with the information in Table~\ref{tbl:songpreds-unknown},
most of the collaborative songs 
in Table~\ref{tbl:songpreds-collab}
were inferred to be mostly matching the style of Lennon.
While four songs were inferred to be written more in McCartney's style,
two exceptions are worth noting.
The songs ``Baby's in Black'' and ``The Word,'' 
according to 
\citet{compton1988mccartney}, 
were both entirely collaborative, with Lennon
having claimed that ``The Word'' was mostly his work.
It is curious, in particular, that ``The Word'' is 
inferred with near certainty of being McCartney-authored.
One feature of the song is the predominance of the flat third.
This McCartney-like motif may be responsible for the high
probability that the song is inferred to be written by
McCartney.
The other two songs, ``From Me to You'' and ``She Loves You,''
were also more likely to be McCartney-authored.
\citet{compton1988mccartney} reported that the former was claimed to
be entirely collaborative, and that the latter was initiated by McCartney
even though the song was written collaboratively.

Two of the collaborations are worthy of comment.
While Lennon and McCartney co-wrote ``She Loves You,'' Lennon remembered that 
``it was Paul's idea'' \citep{compton1988mccartney},
and the probability indicates that the song is weighted towards McCartney. 
On the other hand, our model's probability prediction for 
``I Want to Hold Your Hand,'' which was written ``eyeball to eyeball'' 
\citep{compton1988mccartney},
is that the song is much more characteristic of Lennon's style. 
Indeed, in one of the Jann Wenner interviews \citep{wenner2009}, 
Lennon opined about 
the beauty of the song's melody, and picked out that song along with his 
song ``Help!'' as the two Beatles' songs he might have wanted to re-record. 
However, perhaps the song might have been special to him as it had much 
more of his imprint. 

Of all Lennon-McCartney songs,
``In My Life'' has probably garnered the greatest amount of speculation
about its true author.
{\em Rolling Stone} magazine considered it to be the 23rd greatest song of
all time ({\em Rolling Stone}, 2011).  
Our model produces a probability of 18.9\% that McCartney
wrote the verse, and a 43.5\%
probability that McCartney wrote the bridge,
with a large amount of uncertainty about the latter.
Because it is known that Lennon wrote the lyrics, it would not
be surprising that he also wrote the music.
Lennon claimed 
\citep{compton1988mccartney} 
that McCartney helped with the bridge, but that was the extent
of his contribution.
Breaking apart the song into the verse and the bridge separately,
it is apparent that the verse is much more consistent stylistically
with Lennon's songwriting.
Thus, a conclusion by our model is that the verse is 
consistent with Lennon's songwriting style, but the bridge
less so.
The bridge having a probability 
that McCartney wrote the song closer to 0.5 may be 
indicative of their collaborative nature, 
as suggested by Lennon, of this part of the song.

\section{Discussion} \label{sec:discuss}

The approach to authorship attribution for Lennon-McCartney
songs we developed in this paper
has connections to methods used 
in attribution analysis of text documents.
One important difference is that typical text analysis models
rely on the relative frequencies of occurrence of words
or word combinations.
In a musical context, where repeats of musical features are intrinsic
to a song's construction, the relative frequencies of the occurrence
of the musical ``words'' may obscure their importance
in characterizing an author's composition style.
Another difference from typical text analysis problems is that
songs include more than just one text stream.
For our work, we specifically included songs' melodic note
sequence and chord sequence as two streams in parallel.
Our particular choice in the representation and
analysis of Lennon-McCartney songs
of the early Beatles period seemed to be sufficient in recovering
a song's author with greater than 75\% accuracy, and with
a high level of discrimination ($c$-statistic of 
0.837 from the ROC analysis).

Our model predictions, particularly for the songs with disputed
authorship, seem to be supported generally with the 
stories that accompany the songs' origins.
While it is tempting to interpret the results of our model
as revelations of a song's true author, other interpretations
are just as compelling.
For example, a disputed song such as ``In My Life'' which according
to our model has a high probability of the verse and bridge each
being written by Lennon,
may in fact have been written by McCartney who stated he composed the song 
in the style of Smokey Robinson and the Miracles \citep{turner1999hard}, 
but actually 
wrote in the style of Lennon, whether consciously or unconsciously.
Songs with high probabilities of being written by Lennon or McCartney are
mainly indications that the songs have musical features that
are consistent with the Lennon or McCartney songs used in the 
development of our model.
To this end, one use of our model is to investigate 
whether certain sections  of disputed or collaborative songs 
are suspected of 
being more consistent with particular composition styles.
For example, the song of disputed authorship ``Wait,'' which our model
estimates a probability of 0.391 of being written by McCartney, 
is sung in harmony by Lennon and McCartney throughout the song
except in the bridge section where McCartney sings alone.
It is natural to ask whether that section may be more in the style
of McCartney who may have had 
a freer hand in writing that portion of the song.
Indeed, our model applied to just the bridge section results in a 
0.646 probability of McCartney authorship, suggesting that the bridge is
more in the style of McCartney than Lennon.

In typical text analyses, the choice of ``stop'' words, i.e.,
the ones used in analyses to distinguish authorship style, 
is often made subjectively or at least by convention.
The analogous decision in a musical context is arguably much
more difficult, as the complexity of choices is far greater.
In our work, we needed to make many subjective decisions that
influenced the construction of musical features.
Such decisions included what constituted
the beginning and ending of melodic phrases,
whether a key change (modulation) should reset the tonic of the song,
whether ad-libbed vocals should be considered part of the melodic line, 
how to include dual melody lines that were sung in harmony,
and so on.
Our guiding principle was to make
choices that could be viewed as the most conservative
in the sense of having the least impact on the information in the data.
For example, we omitted melodic information from ad-libbed vocals,
and made phrasings of melodic lines as long as possible, as shorter
lines introduced extra ``rests'' as part of the melodic transitions.
Also, when it was not clear in cases of dual melody lines which was
the main melody, we included both melody lines.

It is worth noting that the model developed here was not
our first attempt.
We explored variations of the presented approach before
arriving at our final model,
including versions that permitted interactions,
alternative variable selection procedures 
such as recursive feature elimination and stepwise variable selection,
models for the musical features as a function of authorship
that were inverted using Bayes rule, random forests,
as well as several others.
A danger in exploring too many models, especially with our
small sample size
and without a true test/holdout set,
is the potential to overfit.
This concern may not be apparent in the presentation of our
analytic summaries, which was the culmination of a series
of model investigations.
The concern of overfitting limited some of our explorations.
For example, after having modest success using elastic net logistic 
regression without any variable pre-processing,
we inserted variable screening parameterized by a $p$-value threshold
based only on four threshold values.
Using a greater range of thresholds,
especially after having learned that elastic net alone was a promising approach,
and that we were tuning the model parameters based on the same 
leave-one-out validation data,
would have had the potential to produce overfitted predictions.
We suspect that our final model, however,
does not suffer from overfitting concerns
in any appreciable way.
First, the approach we present is actually
fairly simple: the removal of musical features
based on bivariate relationships with the response
followed by regularized logistic regression.
More complex procedures might raise questions about their
generalizability.
Second, we were cautious about optimizing the prediction algorithm
and calibrating the predictability
using out-of-sample criteria.
For example, probability predictions involved 
leaving out data (one song at a time) to optimize the $p$-value threshold
for variable screening, followed by leaving out portions of data 
(20\% of the data that remained) to optimize the elastic net tuning
parameters; and this entire procedure was performed leaving out
one song at a time when making predictions for the songs of known
authorship.
This cascading application of cross-validation mitigates 
some of the natural concerns about possible overfitting.

Our particular modeling approach does permit 
extensions to address wider sets of songwriter 
attribution applications.
Our model assumes only two authors,
but this is easily extended to multiple songwriters in 
larger applications by modeling authorship in a multinomial logit
model, for example.
Another extension of our model can address changes in an 
author's style over time.
Our application to Lennon-McCartney songs focused
on a time period where the songwriters' musical styles were not changing
in profound ways.
To include larger spans of time where a songwriter's style may
be changing, one possibility is to assume a stochastic process
on the musical feature effects for each author, 
such as through an autoregressive process.
Such an approach acknowledges that an author's style
is likely to evolve gradually over time and with an uncertain trajectory.
This approach would be straightforward to implement in a Bayesian
setting, though implementing such a model
in conjunction with variable screening
would involve methodological challenges.

Several other limitations are worth mentioning.
Our approach assumes that each song or (more relevantly) song portion
contains sufficiently rich detail to capture musical information 
for distinguishing authorship.
Shorter song fragments would have a scarcity of features, 
and probability predictions are expected to be less reliable.
Furthermore,
if the goal of this work was to make the most accurate predictions
of a song's author, then our approach could clearly be improved by
incorporating readily available additional information.
Lyric content, information on a song's structure, 
use of rhythm, song tempo, time signature, and 
the identity of a song's actual singer or singers are all
likely to be highly predictive and distinguishing of a song's
authorship.
Our decision to ignore this extra information is consistent
with the larger goal of being able to establish the stylistic
fingerprint of a songwriter based solely on a corpus of songs'
musical content, 
using Lennon-McCartney songs as a sandbox for 
understanding the potential for this approach.
Ultimately, the reduction of a songwriter's musical content 
into low-dimensional representations, such as a vector of
musical feature effects, is the first step 
towards establishing musical signatures for songwriters
that can be used for further analysis.
For example, with many songwriters' styles characterized
in a reduced form, 
it is possible to establish influence networks to learn about
the diffusion of the creative process in popular music.
With recent improvements in technology to convert 
audio information into formats amenable to the type of
analysis we developed in this paper
\citep{casey2008content,fu2011survey},
larger-scale analyses of songwriters' styles are
a potential area of exploration.

\appendix
\section{Musical Background}  \label{sec:musicbg} 

\begin{figure}[ht]
\centering
\includegraphics[width=1.0\textwidth]{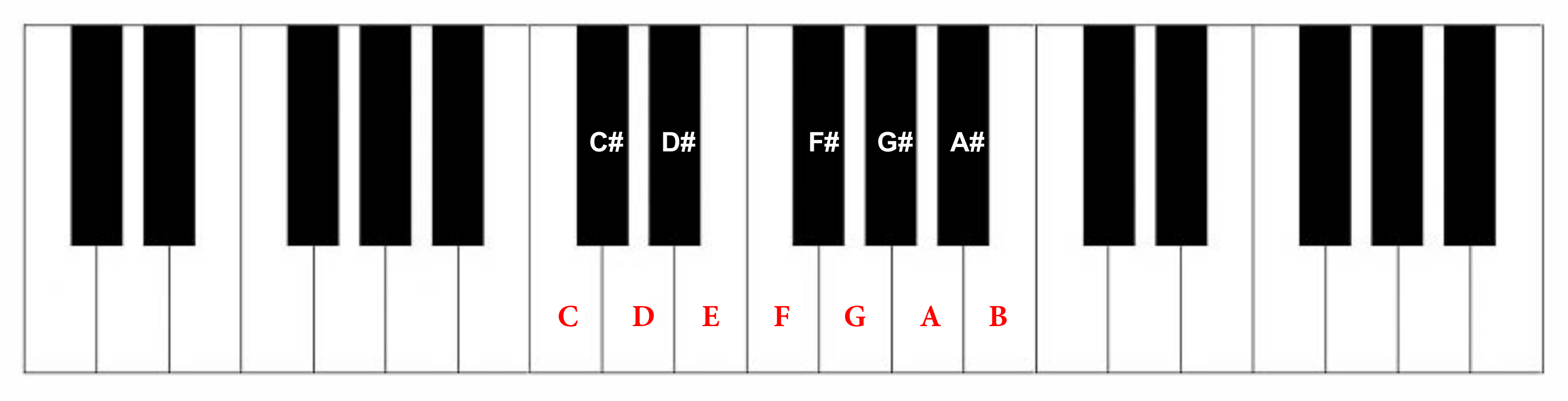}
\caption{\label{fig:chromatic}
Chromatic scale notes appearing on a piano diagram.
}
\end{figure}

A justification for the musical features chosen requires an understanding of Western popular music. 
{\em Middle C}, often denoted as C4, has frequency $261.6$Hz, and the well known 
equally-tempered 12-tone {\em chromatic scale} starting on note C4 is the sequence of notes
\begin{center}
C4, C\#4, D4, D\#4, E4, F4, F\#4, G4, G\#4, A4, A\#4, B4 
\end{center}
where each successive note is derived from the previous one by multiplying the 
frequency by $2^{1/12}$. 
In the above sequence, notes preceding the ``4'' 
(i.e., C, C\#, D, D\#, E, F, F\#, G, G\#, A, A\#, B)
are the {\em pitches}, and the number 4 refers to the {\em octave} of the note.
The continuation of the sequence above is the same set of pitches, but at
the next higher octave, that is, C5, C\#5, D5, and so on.
The 12 notes can also be visualized in a piano diagram in Figure~\ref{fig:chromatic}.

For the current discussion,
we can represent
a note as $Z[i,j]$, where $i\geq 1$ indexes the pitch of the note
and $j\geq 1$ indexes the octave of the note.
We set $Z[1,4]= \mbox{C4}$, and all other notes are relative to this anchoring choice.
Given the circular ordering of pitches in the chromatic scale,
$Z[i+12,j] = Z[i,j+1]$ 
for all $i$ and $j$.
Thus, a specific note has multiple representations using this notation.
By convention, the octave of a note is the value $j$ in which
the representation $Z[i,j]$ has $i\leq 12$.

The notes $Z[i,j]$ and $Z[i+1,j]$ are said to be a
{\em semitone apart}, while the notes 
$Z[i,j]$ and $Z[i+2,j]$
are said to be a {\em whole tone} apart. 
Notes $Z[i,j], Z[i,j+1], Z[i,j+2],\ldots$, 
are said to be in the same {\em pitch class}.
Thus, 
D3, D4, D5, 
and so on, are in the same pitch class,
but reside in different octaves.
It is worth noting 
that while the {\em sharp} symbol $\#$ denotes raising a note a semitone, 
one can also use the {\em flat} suffix $\flat$ to lower a note a semitone. 
One can translate or {\em transpose} the chromatic scale to start on any note
given its circular structure, and to the human ear all such chromatic scales 
played in sequence sound  essentially the same. 
A chromatic scale can start on any note $Z[i,j]$ and consists of the 12 notes
$(Z[i,j], Z[i+1,j], \ldots, Z[i+11,j])$.

The basis of Western music is the {\em diatonic scale}, which, 
starting on a given note $Z[i,j]$, 
called the {\em tonic} of the scale,
consists of the subsequence of seven notes from the chromatic scale
\[
(Z[i,j], Z[i+2,j], Z[i+4,j], Z[i+5,j], Z[i+7,j], Z[i+9,j], Z[i+11,j]).
\]
For example, beginning on an A at any octave, 
the diatonic scale with tonic A is 
(A, B, C\#, D, E, F\#, G\#).
Chromatic notes that are not part of the diatonic scale are called {\em non-diatonic}. 
Thus the non-diatonic notes with respect to the diatonic scale starting on A
include 
A\#, C, D\#, F, and G.

The diatonic scale permeates much of Western music, and most popular songs (or portions of songs) 
can be analyzed to be based on a diatonic scale starting at a specific note
belonging to one of the 12 pitch classes; the lowest note of the diatonic scale is called 
the {\em major key}, or just the {\em key}, of the song, and the note itself is the tonic of that key.
Songs are often to be found in a ``minor'' key, based on a {\em minor scale}. 
For our purposes, we associate, as is often done, the minor key with the 
major key three semitones up, 
as they share the same seven notes.
This particular definition of a minor key is often called the 
{\em natural minor}, and is the
{\em relative minor} of the associated major key. 
For example, the key of A minor (as a natural minor) 
consists of the notes 
(A, B, C, D, E, F, G), 
which
are the same as those in the major key of C 
(C, D, E, F, G, A, B), 
so that A minor is the 
relative minor associated with C major.
Because the major key and relative minor share the same notes on the diatonic scale,
in our work we classify songs being in the major key as a proxy for the diatonic notes.

With a given key of a song,
non-diatonic notes are usually specified by their relation to the tonic.
So, for example, in the key of C, the flat third and flat seventh are E$\flat$ and B$\flat$ 
(and they could, equivalently, be called the raised second and raised sixth, as well). 
In fact, in pop/rock music, the flat third and flat seventh play a large role, 
as they appear in the five note  {\em pentatonic} (or the {\em blues}) scale, 
which consists of the notes $(Z[i,j], Z[i+3,j], Z[i+5,j], Z[i+7,j], Z[i+10,j])$, where
$Z[i,j]$ is the tonic of the pentatonic scale.
Thus, the pentatonic scale starting on tonic C is 
(C, E$\flat$, F, G, B$\flat$).

A {\em note transition} or an {\em interval} is a pair of notes, where the {\em size} of the interval 
depends on the number of semitones between them.
Some sample intervals include:
\begin{itemize}
\item {\em unison} is between two identical notes (e.g., 
C4 $\rightarrow$ C4).
\item a {\em major second} consists of two notes where the second is two semitones 
(whole tone) up from the first (e.g., 
C4 $\rightarrow$ D4,
F4 $\rightarrow$ G4).
\item a {\em major third} consists of two notes where the second is four semitones 
(two whole tones) up from the first (e.g., 
C4 $\rightarrow$ E4,
F4 $\rightarrow$ A4).
\item a {\em perfect fourth} consists of two notes where the second is five semitones 
up from the first (e.g., 
D4 $\rightarrow$ G4).
\item a {\em perfect fifth} consists of two notes where the second is seven semitones 
up from the first (e.g., 
A4 $\rightarrow$ E5).
\item a {\em major sixth} consists of two notes where the second is nine semitones 
up from the first (e.g., 
D4 $\rightarrow$ B4).
\item a {\em major seventh} consists of two notes where the second is 11 semitones 
up from the first (e.g., 
F4 $\rightarrow$ E5).
\item an {\em octave} consists of two notes where the second is 12 semitones up 
from the first (e.g., 
C4 $\rightarrow$ C5).
\end{itemize}
The minor second, third, sixth, and seventh intervals arise by lowering the 
second note of the 
corresponding major interval by a semitone. 
For example, 
C $\rightarrow$  E$\flat$ 
is a minor third. 
For intervals of a fourth and fifth, the term {\em diminished} applies when the top note 
of the corresponding interval is decreased by a semitone,
and the term {\em augmented} applies when raising the top note a semitone.
As an example,
the interval 
C $\rightarrow$ G\# 
is an augmented fifth
in the key of C. 
In our choice of note transitions within pop songs, the diatonic notes (always relative to the key) have 
prime importance, with special emphasis on diatonic transitions to and from the tonic, transitions between 
small steps on the diatonic scale (which are fairly common in melody writing), and transitions 
along the pentatonic/blues scale.

Chords, for our purposes, consist of three notes played simultaneously (called a {\em triad}), 
and form the basis of most of the harmony in pop songs. 
The two most common types of chords are {\em major} chords and {\em minor} chords.
A major chord is formed, using $Z[i,j]$ as the root of the chord, as 
$(Z[i,j], Z[i+4,j], Z[i+7,j])$.
A minor chord, in contrast, is formed as 
$(Z[i,j], Z[i+3,j], Z[i+7,j])$.
Less common are diminished chords, formed as
$(Z[i,j], Z[i+3,j], Z[i+6,j])$, and
augmented chords, formed as
$(Z[i,j], Z[i+4,j], Z[i+8,j])$.
Building chords from the diatonic scale consists of taking a starting note within the scale 
and successively layering on two extra notes above it, skipping a note each time. 
For example, in the key of C, the {\em diatonic chords} are:
\begin{itemize}
\item C major, the I major chord
(the {\em tonic}), consisting of notes C, E, and G.
\item D minor, the ii minor
chord, consisting of notes D, F, and A.
\item E minor, the iii minor
chord, consisting of notes E, G, and B.
\item F major, the IV major
chord (the {\em subdominant}), consisting of notes F, A, and C.
\item G major, the V major chord
(the {\em dominant}), consisting of notes G, B, and D.
\item A minor, the vi minor
chord, consisting of notes A, C, and E.
\item B diminished, the
vii$^\circ$ diminished chord, consisting of notes B, D, and F.
\end{itemize}
All of these diatonic chords are ``native'' to the scale in which they reside; 
all other chords, with respect to the scale, are deemed to be {\em non-diatonic chords}. 
The diatonic chords are the most common ones in popular songs, although non-diatonic chords 
are often added for variety and creating emotional tension. 
In particular, in rock-and-roll music, the major chords on the flat third and the flat seventh 
(and sometimes the flat sixth) play a significant role in that genre.

In pop/rock music, the diatonic chords are all prevalent, especially the tonic (I), 
subdominant (IV), and dominant (V) chords, with the exception of the diminished seventh chord 
on the seventh note  of the diatonic scale; this chord is rarely used.
The minor chord on the seventh note occurs more often,
and is sometimes considered a replacement as one of the diatonic chords.

Transitions between chords are a cornerstone of pop/rock music.
{\em Chord progressions} are sequences of chords that often repeat
throughout a song.
Transitions between diatonic chords form the bulk of the chord transitions.
Less common (but not infrequently, when grouped together) are transitions 
between non-diatonic chords and the tonic (I) or dominant (V).

Entire songs can be viewed in their most basic form
as the superposition of chord progressions along with 
melodic lines.
Songs are divided into sections within which chord progressions
and melodies are identical or nearly identical.
Two of the main sections that appear in most pop/rock songs
are the verse and the chorus.
The {\em verses} within a song typically have identical musical content,
but usually contain different lyrics.
The {\em chorus} of a song typically has greater musical and emotional
intensity than the verse, 
and contains identical lyrics across repeats within the song.
It is common for songs to have a third musical section inserted between
an occurrence of the chorus and a subsequent verse, called the {\em bridge} section.
This section musically functions as a connector between the chorus and verse,
and may even undergo a {\em modulation}, that is, resetting the song
to a different key, if only temporarily.
Other types of sections may appear in typical pop/rock music
(e.g., intro, pre-chorus, outro), but the verse, chorus, and bridge are
nearly universal components of a song.

More details about the basics of melodic and harmonic structure of popular
music can be found in 
\citet{benward2014music}
and 
\citet{middleton1990studying}.

\nocite{rolling2011}

\Urlmuskip=0mu plus 1mu\relax
\singlespacing
\bibliographystyle{apacite}
\setlength\bibsep{0.75\baselineskip}
\bibliography{beatles}

\begin{thebibliography}{}

\bibitem [\protect \citeauthoryear {%
Airoldi%
, Anderson%
, Fienberg%
\BCBL {}\ \BBA {} Skinner%
}{%
Airoldi%
\ \protect \BOthers {.}}{%
{\protect \APACyear {2006}}%
}]{%
airoldi2006wrote}
\APACinsertmetastar {%
airoldi2006wrote}%
\begin{APACrefauthors}%
Airoldi, E\BPBI M.%
, Anderson, A\BPBI G.%
, Fienberg, S\BPBI E.%
\BCBL {}\ \BBA {} Skinner, K\BPBI K.%
\end{APACrefauthors}%
\unskip\
\newblock
\APACrefYearMonthDay{2006}{}{}.
\newblock
{\BBOQ}\APACrefatitle {Who wrote {R}onald {R}eagan's radio addresses?} {Who
  wrote {R}onald {R}eagan's radio addresses?}{\BBCQ}
\newblock
\APACjournalVolNumPages{Bayesian Analysis}{1}{2}{289--319}.
\PrintBackRefs{\CurrentBib}

\bibitem [\protect \citeauthoryear {%
Benward%
}{%
Benward%
}{%
{\protect \APACyear {2014}}%
}]{%
benward2014music}
\APACinsertmetastar {%
benward2014music}%
\begin{APACrefauthors}%
Benward, B.%
\end{APACrefauthors}%
\unskip\
\newblock
\APACrefYear{2014}.
\newblock
\APACrefbtitle {Music in Theory and Practice, Volume 1} {Music in theory and
  practice, volume 1}.
\newblock
\APACaddressPublisher{}{McGraw-Hill Higher Education}.
\PrintBackRefs{\CurrentBib}

\bibitem [\protect \citeauthoryear {%
Bien%
, Taylor%
\BCBL {}\ \BBA {} Tibshirani%
}{%
Bien%
\ \protect \BOthers {.}}{%
{\protect \APACyear {2013}}%
}]{%
bien2013lasso}
\APACinsertmetastar {%
bien2013lasso}%
\begin{APACrefauthors}%
Bien, J.%
, Taylor, J.%
\BCBL {}\ \BBA {} Tibshirani, R.%
\end{APACrefauthors}%
\unskip\
\newblock
\APACrefYearMonthDay{2013}{}{}.
\newblock
{\BBOQ}\APACrefatitle {A lasso for hierarchical interactions} {A lasso for
  hierarchical interactions}.{\BBCQ}
\newblock
\APACjournalVolNumPages{The {A}nnals of {S}tatistics}{41}{3}{1111-1141}.
\PrintBackRefs{\CurrentBib}

\bibitem [\protect \citeauthoryear {%
Breiman%
}{%
Breiman%
}{%
{\protect \APACyear {2001}}%
}]{%
breiman2001random}
\APACinsertmetastar {%
breiman2001random}%
\begin{APACrefauthors}%
Breiman, L.%
\end{APACrefauthors}%
\unskip\
\newblock
\APACrefYearMonthDay{2001}{}{}.
\newblock
{\BBOQ}\APACrefatitle {Random forests} {Random forests}.{\BBCQ}
\newblock
\APACjournalVolNumPages{Machine {L}earning}{45}{1}{5--32}.
\PrintBackRefs{\CurrentBib}

\bibitem [\protect \citeauthoryear {%
Brown%
}{%
Brown%
}{%
{\protect \APACyear {2004}}%
}]{%
brown2004mathematics}
\APACinsertmetastar {%
brown2004mathematics}%
\begin{APACrefauthors}%
Brown, J\BPBI I.%
\end{APACrefauthors}%
\unskip\
\newblock
\APACrefYearMonthDay{2004}{}{}.
\newblock
{\BBOQ}\APACrefatitle {Mathematics, Physics and {A Hard Day's Night}}
  {Mathematics, physics and {A Hard Day's Night}}.{\BBCQ}
\newblock
\APACjournalVolNumPages{CMS Notes}{36}{6}{4--8}.
\PrintBackRefs{\CurrentBib}

\bibitem [\protect \citeauthoryear {%
Casey%
\ \protect \BOthers {.}}{%
Casey%
\ \protect \BOthers {.}}{%
{\protect \APACyear {2008}}%
}]{%
casey2008content}
\APACinsertmetastar {%
casey2008content}%
\begin{APACrefauthors}%
Casey, M\BPBI A.%
, Veltkamp, R.%
, Goto, M.%
, Leman, M.%
, Rhodes, C.%
\BCBL {}\ \BBA {} Slaney, M.%
\end{APACrefauthors}%
\unskip\
\newblock
\APACrefYearMonthDay{2008}{}{}.
\newblock
{\BBOQ}\APACrefatitle {Content-based music information retrieval: Current
  directions and future challenges} {Content-based music information retrieval:
  Current directions and future challenges}.{\BBCQ}
\newblock
\APACjournalVolNumPages{Proceedings of the IEEE}{96}{4}{668--696}.
\PrintBackRefs{\CurrentBib}

\bibitem [\protect \citeauthoryear {%
Cath{\'e}%
}{%
Cath{\'e}%
}{%
{\protect \APACyear {2016}}%
}]{%
cathe2016nostalgie}
\APACinsertmetastar {%
cathe2016nostalgie}%
\begin{APACrefauthors}%
Cath{\'e}, P.%
\end{APACrefauthors}%
\unskip\
\newblock
\APACrefYearMonthDay{2016}{}{}.
\newblock
{\BBOQ}\APACrefatitle {La nostalgie chez les {B}eatles: vers une application de
  la th{\'e}orie des vecteurs harmoniques {\`a} la musique pop?} {La nostalgie
  chez les {B}eatles: vers une application de la th{\'e}orie des vecteurs
  harmoniques {\`a} la musique pop?}{\BBCQ}
\newblock
\APACjournalVolNumPages{Volume!}{12}{1}{181--191}.
\PrintBackRefs{\CurrentBib}

\bibitem [\protect \citeauthoryear {%
Chipman%
, George%
\BCBL {}\ \BBA {} McCulloch%
}{%
Chipman%
\ \protect \BOthers {.}}{%
{\protect \APACyear {1998}}%
}]{%
chipman1998bayesian}
\APACinsertmetastar {%
chipman1998bayesian}%
\begin{APACrefauthors}%
Chipman, H\BPBI A.%
, George, E\BPBI I.%
\BCBL {}\ \BBA {} McCulloch, R\BPBI E.%
\end{APACrefauthors}%
\unskip\
\newblock
\APACrefYearMonthDay{1998}{}{}.
\newblock
{\BBOQ}\APACrefatitle {Bayesian {CART} model search} {Bayesian {CART} model
  search}.{\BBCQ}
\newblock
\APACjournalVolNumPages{Journal of the American Statistical
  Association}{93}{443}{935--948}.
\PrintBackRefs{\CurrentBib}

\bibitem [\protect \citeauthoryear {%
Cilibrasi%
, Vit{\'a}nyi%
\BCBL {}\ \BBA {} De~Wolf%
}{%
Cilibrasi%
\ \protect \BOthers {.}}{%
{\protect \APACyear {2004}}%
}]{%
cilibrasi2004algorithmic}
\APACinsertmetastar {%
cilibrasi2004algorithmic}%
\begin{APACrefauthors}%
Cilibrasi, R.%
, Vit{\'a}nyi, P.%
\BCBL {}\ \BBA {} De~Wolf, R.%
\end{APACrefauthors}%
\unskip\
\newblock
\APACrefYearMonthDay{2004}{}{}.
\newblock
{\BBOQ}\APACrefatitle {Algorithmic clustering of music based on string
  compression} {Algorithmic clustering of music based on string
  compression}.{\BBCQ}
\newblock
\APACjournalVolNumPages{Computer Music Journal}{28}{4}{49--67}.
\PrintBackRefs{\CurrentBib}

\bibitem [\protect \citeauthoryear {%
Clement%
\ \BBA {} Sharp%
}{%
Clement%
\ \BBA {} Sharp%
}{%
{\protect \APACyear {2003}}%
}]{%
clement2003ngram}
\APACinsertmetastar {%
clement2003ngram}%
\begin{APACrefauthors}%
Clement, R.%
\BCBT {}\ \BBA {} Sharp, D.%
\end{APACrefauthors}%
\unskip\
\newblock
\APACrefYearMonthDay{2003}{}{}.
\newblock
{\BBOQ}\APACrefatitle {{\em N}-gram and {B}ayesian classification of documents
  for topic and authorship} {{\em N}-gram and {B}ayesian classification of
  documents for topic and authorship}.{\BBCQ}
\newblock
\APACjournalVolNumPages{Literary and {L}inguistic
  {C}omputing}{18}{4}{423--447}.
\PrintBackRefs{\CurrentBib}

\bibitem [\protect \citeauthoryear {%
Compton%
}{%
Compton%
}{%
{\protect \APACyear {1988}}%
}]{%
compton1988mccartney}
\APACinsertmetastar {%
compton1988mccartney}%
\begin{APACrefauthors}%
Compton, T.%
\end{APACrefauthors}%
\unskip\
\newblock
\APACrefYearMonthDay{1988}{}{}.
\newblock
{\BBOQ}\APACrefatitle {{M}c{C}artney or {L}ennon?: {B}eatles Myths and the
  Composing of the {L}ennon-{M}c{C}artney Songs} {{M}c{C}artney or {L}ennon?:
  {B}eatles myths and the composing of the {L}ennon-{M}c{C}artney
  songs}.{\BBCQ}
\newblock
\APACjournalVolNumPages{The Journal of Popular Culture}{22}{2}{99--131}.
\PrintBackRefs{\CurrentBib}

\bibitem [\protect \citeauthoryear {%
Conklin%
}{%
Conklin%
}{%
{\protect \APACyear {2006}}%
}]{%
conklin2006melodic}
\APACinsertmetastar {%
conklin2006melodic}%
\begin{APACrefauthors}%
Conklin, D.%
\end{APACrefauthors}%
\unskip\
\newblock
\APACrefYearMonthDay{2006}{}{}.
\newblock
{\BBOQ}\APACrefatitle {Melodic analysis with segment classes} {Melodic analysis
  with segment classes}.{\BBCQ}
\newblock
\APACjournalVolNumPages{Machine Learning}{65}{2}{349--360}.
\PrintBackRefs{\CurrentBib}

\bibitem [\protect \citeauthoryear {%
Draper%
}{%
Draper%
}{%
{\protect \APACyear {2013}}%
}]{%
draper2013bayesian}
\APACinsertmetastar {%
draper2013bayesian}%
\begin{APACrefauthors}%
Draper, D.%
\end{APACrefauthors}%
\unskip\
\newblock
\APACrefYearMonthDay{2013}{}{}.
\newblock
{\BBOQ}\APACrefatitle {Bayesian model specification: {H}euristics and examples}
  {Bayesian model specification: {H}euristics and examples}.{\BBCQ}
\newblock
\BIn{} P.~Damien, P.~Dellaportas, N\BPBI G.~Polson\BCBL {}\ \BBA {} D\BPBI
  A.~Stephens\ (\BEDS), \APACrefbtitle {Bayesian Theory and Applications}
  {Bayesian theory and applications}\ (\BPGS\ 409--431).
\newblock
\APACaddressPublisher{}{New York: Oxford University Press}.
\PrintBackRefs{\CurrentBib}

\bibitem [\protect \citeauthoryear {%
Dubnov%
, Assayag%
, Lartillot%
\BCBL {}\ \BBA {} Bejerano%
}{%
Dubnov%
\ \protect \BOthers {.}}{%
{\protect \APACyear {2003}}%
}]{%
dubnov2003using}
\APACinsertmetastar {%
dubnov2003using}%
\begin{APACrefauthors}%
Dubnov, S.%
, Assayag, G.%
, Lartillot, O.%
\BCBL {}\ \BBA {} Bejerano, G.%
\end{APACrefauthors}%
\unskip\
\newblock
\APACrefYearMonthDay{2003}{}{}.
\newblock
{\BBOQ}\APACrefatitle {Using machine-learning methods for musical style
  modeling} {Using machine-learning methods for musical style modeling}.{\BBCQ}
\newblock
\APACjournalVolNumPages{Computer}{36}{10}{73--80}.
\PrintBackRefs{\CurrentBib}

\bibitem [\protect \citeauthoryear {%
Efron%
\ \BBA {} Thisted%
}{%
Efron%
\ \BBA {} Thisted%
}{%
{\protect \APACyear {1976}}%
}]{%
efron1976estimating}
\APACinsertmetastar {%
efron1976estimating}%
\begin{APACrefauthors}%
Efron, B.%
\BCBT {}\ \BBA {} Thisted, R.%
\end{APACrefauthors}%
\unskip\
\newblock
\APACrefYearMonthDay{1976}{}{}.
\newblock
{\BBOQ}\APACrefatitle {Estimating the number of unseen species: How many words
  did {S}hakespeare know?} {Estimating the number of unseen species: How many
  words did {S}hakespeare know?}{\BBCQ}
\newblock
\APACjournalVolNumPages{Biometrika}{63}{3}{435--447}.
\PrintBackRefs{\CurrentBib}

\bibitem [\protect \citeauthoryear {%
Everett%
}{%
Everett%
}{%
{\protect \APACyear {1999}}%
}]{%
everett1999beatles}
\APACinsertmetastar {%
everett1999beatles}%
\begin{APACrefauthors}%
Everett, W.%
\end{APACrefauthors}%
\unskip\
\newblock
\APACrefYear{1999}.
\newblock
\APACrefbtitle {The {B}eatles as Musicians: Revolver through the Anthology}
  {The {B}eatles as musicians: Revolver through the anthology}.
\newblock
\APACaddressPublisher{}{Oxford University Press, USA}.
\PrintBackRefs{\CurrentBib}

\bibitem [\protect \citeauthoryear {%
Fan%
}{%
Fan%
}{%
{\protect \APACyear {2007}}%
}]{%
fan2007variable}
\APACinsertmetastar {%
fan2007variable}%
\begin{APACrefauthors}%
Fan, J.%
\end{APACrefauthors}%
\unskip\
\newblock
\APACrefYearMonthDay{2007}{}{}.
\newblock
{\BBOQ}\APACrefatitle {Variable screening in high-dimensional feature space}
  {Variable screening in high-dimensional feature space}.{\BBCQ}
\newblock
\BIn{} \APACrefbtitle {Proceedings of the 4th International Congress of Chinese
  Mathematicians} {Proceedings of the 4th international congress of chinese
  mathematicians}\ (\BVOL~2, \BPGS\ 735--747).
\PrintBackRefs{\CurrentBib}

\bibitem [\protect \citeauthoryear {%
Fan%
\ \BBA {} Lv%
}{%
Fan%
\ \BBA {} Lv%
}{%
{\protect \APACyear {2008}}%
}]{%
fan2008sure}
\APACinsertmetastar {%
fan2008sure}%
\begin{APACrefauthors}%
Fan, J.%
\BCBT {}\ \BBA {} Lv, J.%
\end{APACrefauthors}%
\unskip\
\newblock
\APACrefYearMonthDay{2008}{}{}.
\newblock
{\BBOQ}\APACrefatitle {Sure independence screening for ultrahigh dimensional
  feature space} {Sure independence screening for ultrahigh dimensional feature
  space}.{\BBCQ}
\newblock
\APACjournalVolNumPages{Journal of the Royal Statistical Society, Series B
  (Statistical Methodology)}{70}{5}{849--911}.
\PrintBackRefs{\CurrentBib}

\bibitem [\protect \citeauthoryear {%
Fan%
\ \BBA {} Song%
}{%
Fan%
\ \BBA {} Song%
}{%
{\protect \APACyear {2010}}%
}]{%
fan2010sure}
\APACinsertmetastar {%
fan2010sure}%
\begin{APACrefauthors}%
Fan, J.%
\BCBT {}\ \BBA {} Song, R.%
\end{APACrefauthors}%
\unskip\
\newblock
\APACrefYearMonthDay{2010}{}{}.
\newblock
{\BBOQ}\APACrefatitle {Sure independence screening in generalized linear models
  with {NP}-dimensionality} {Sure independence screening in generalized linear
  models with {NP}-dimensionality}.{\BBCQ}
\newblock
\APACjournalVolNumPages{The Annals of Statistics}{38}{6}{3567--3604}.
\PrintBackRefs{\CurrentBib}

\bibitem [\protect \citeauthoryear {%
Friedman%
, Hastie%
\BCBL {}\ \BBA {} Tibshirani%
}{%
Friedman%
\ \protect \BOthers {.}}{%
{\protect \APACyear {2010}}%
}]{%
glmnet2010}
\APACinsertmetastar {%
glmnet2010}%
\begin{APACrefauthors}%
Friedman, J.%
, Hastie, T.%
\BCBL {}\ \BBA {} Tibshirani, R.%
\end{APACrefauthors}%
\unskip\
\newblock
\APACrefYearMonthDay{2010}{}{}.
\newblock
{\BBOQ}\APACrefatitle {Regularization Paths for Generalized Linear Models via
  Coordinate Descent} {Regularization paths for generalized linear models via
  coordinate descent}.{\BBCQ}
\newblock
\APACjournalVolNumPages{Journal of Statistical Software}{33}{1}{}.
\newblock
\begin{APACrefURL} \url{http://www.jstatsoft.org/v33/i01/} \end{APACrefURL}
\PrintBackRefs{\CurrentBib}

\bibitem [\protect \citeauthoryear {%
Fu%
, Lu%
, Ting%
\BCBL {}\ \BBA {} Zhang%
}{%
Fu%
\ \protect \BOthers {.}}{%
{\protect \APACyear {2011}}%
}]{%
fu2011survey}
\APACinsertmetastar {%
fu2011survey}%
\begin{APACrefauthors}%
Fu, Z.%
, Lu, G.%
, Ting, K\BPBI M.%
\BCBL {}\ \BBA {} Zhang, D.%
\end{APACrefauthors}%
\unskip\
\newblock
\APACrefYearMonthDay{2011}{}{}.
\newblock
{\BBOQ}\APACrefatitle {A survey of audio-based music classification and
  annotation} {A survey of audio-based music classification and
  annotation}.{\BBCQ}
\newblock
\APACjournalVolNumPages{IEEE {T}ransactions on {M}ultimedia}{13}{2}{303--319}.
\PrintBackRefs{\CurrentBib}

\bibitem [\protect \citeauthoryear {%
Fujita%
, Hagino%
, Kubo%
\BCBL {}\ \BBA {} Sato%
}{%
Fujita%
\ \protect \BOthers {.}}{%
{\protect \APACyear {1993}}%
}]{%
fujita1993beatles}
\APACinsertmetastar {%
fujita1993beatles}%
\begin{APACrefauthors}%
Fujita, T.%
, Hagino, Y.%
, Kubo, H.%
\BCBL {}\ \BBA {} Sato, G.%
\end{APACrefauthors}%
\unskip\
\newblock
\APACrefYear{1993}.
\newblock
\APACrefbtitle {The {B}eatles: Complete {S}cores} {The {B}eatles: Complete
  {S}cores}.
\newblock
\APACaddressPublisher{}{Hal Leonard Publishing Corporation}.
\PrintBackRefs{\CurrentBib}

\bibitem [\protect \citeauthoryear {%
George%
\ \BBA {} Shamir%
}{%
George%
\ \BBA {} Shamir%
}{%
{\protect \APACyear {2014}}%
}]{%
george2014computer}
\APACinsertmetastar {%
george2014computer}%
\begin{APACrefauthors}%
George, J.%
\BCBT {}\ \BBA {} Shamir, L.%
\end{APACrefauthors}%
\unskip\
\newblock
\APACrefYearMonthDay{2014}{}{}.
\newblock
{\BBOQ}\APACrefatitle {Computer analysis of similarities between albums in
  popular music} {Computer analysis of similarities between albums in popular
  music}.{\BBCQ}
\newblock
\APACjournalVolNumPages{Pattern Recognition Letters}{45}{}{78--84}.
\PrintBackRefs{\CurrentBib}

\bibitem [\protect \citeauthoryear {%
Hartzog%
}{%
Hartzog%
}{%
{\protect \APACyear {2016}}%
}]{%
hartzog2016}
\APACinsertmetastar {%
hartzog2016}%
\begin{APACrefauthors}%
Hartzog, B.%
\end{APACrefauthors}%
\unskip\
\newblock
\APACrefYearMonthDay{2016}{March}{}.
\newblock
\APACrefbtitle {The {B}eatles' Songwriting.} {The {B}eatles' songwriting.}
\newblock
\begin{APACrefURL}
  \url{http://www.brianhartzog.com/beatles/beatles-songwriting.htm}
  \end{APACrefURL}
\newblock
\APACrefnote{Accessed 07-June-2017}
\PrintBackRefs{\CurrentBib}

\bibitem [\protect \citeauthoryear {%
Heuger%
}{%
Heuger%
}{%
{\protect \APACyear {2018}}%
}]{%
heuger2018}
\APACinsertmetastar {%
heuger2018}%
\begin{APACrefauthors}%
Heuger, M.%
\end{APACrefauthors}%
\unskip\
\newblock
\APACrefYearMonthDay{2018}{}{}.
\newblock
\APACrefbtitle {Beabliography: Mostly Academic Writings about the {B}eatles.}
  {Beabliography: Mostly academic writings about the {B}eatles.}
\newblock
\begin{APACrefURL} \url{http://www.icce.rug.nl/~soundscapes/BEAB/index.shtml}
  \end{APACrefURL}
\newblock
\APACrefnote{Accessed 11-July-2018}
\PrintBackRefs{\CurrentBib}

\bibitem [\protect \citeauthoryear {%
Hope%
}{%
Hope%
}{%
{\protect \APACyear {1968}}%
}]{%
hope1968simplified}
\APACinsertmetastar {%
hope1968simplified}%
\begin{APACrefauthors}%
Hope, A\BPBI C.%
\end{APACrefauthors}%
\unskip\
\newblock
\APACrefYearMonthDay{1968}{}{}.
\newblock
{\BBOQ}\APACrefatitle {A Simplified {M}onte {C}arlo Significance Test
  Procedure} {A simplified {M}onte {C}arlo significance test procedure}.{\BBCQ}
\newblock
\APACjournalVolNumPages{Journal of the Royal Statistical Society, Series B
  (Statistical Methodology)}{30}{3}{582--598}.
\PrintBackRefs{\CurrentBib}

\bibitem [\protect \citeauthoryear {%
Kempfert%
\ \BBA {} Wong%
}{%
Kempfert%
\ \BBA {} Wong%
}{%
{\protect \APACyear {2018}}%
}]{%
kempfert2018does}
\APACinsertmetastar {%
kempfert2018does}%
\begin{APACrefauthors}%
Kempfert, K\BPBI C.%
\BCBT {}\ \BBA {} Wong, S\BPBI W.%
\end{APACrefauthors}%
\unskip\
\newblock
\APACrefYearMonthDay{2018}{}{}.
\newblock
{\BBOQ}\APACrefatitle {Where Does {H}aydn End and {M}ozart Begin? {C}omposer
  Classification of String Quartets} {Where does {H}aydn end and {M}ozart
  begin? {C}omposer classification of string quartets}.{\BBCQ}
\newblock
\APACjournalVolNumPages{arXiv preprint arXiv:1809.05075}{}{}{}.
\PrintBackRefs{\CurrentBib}

\bibitem [\protect \citeauthoryear {%
Le~Cessie%
\ \BBA {} Van~Houwelingen%
}{%
Le~Cessie%
\ \BBA {} Van~Houwelingen%
}{%
{\protect \APACyear {1992}}%
}]{%
le1992ridge}
\APACinsertmetastar {%
le1992ridge}%
\begin{APACrefauthors}%
Le~Cessie, S.%
\BCBT {}\ \BBA {} Van~Houwelingen, J\BPBI C.%
\end{APACrefauthors}%
\unskip\
\newblock
\APACrefYearMonthDay{1992}{}{}.
\newblock
{\BBOQ}\APACrefatitle {Ridge estimators in logistic regression} {Ridge
  estimators in logistic regression}.{\BBCQ}
\newblock
\APACjournalVolNumPages{Applied Statistics}{41}{1}{191--201}.
\PrintBackRefs{\CurrentBib}

\bibitem [\protect \citeauthoryear {%
Lim%
\ \BBA {} Hastie%
}{%
Lim%
\ \BBA {} Hastie%
}{%
{\protect \APACyear {2015}}%
}]{%
lim2015learning}
\APACinsertmetastar {%
lim2015learning}%
\begin{APACrefauthors}%
Lim, M.%
\BCBT {}\ \BBA {} Hastie, T.%
\end{APACrefauthors}%
\unskip\
\newblock
\APACrefYearMonthDay{2015}{}{}.
\newblock
{\BBOQ}\APACrefatitle {Learning interactions via hierarchical group-lasso
  regularization} {Learning interactions via hierarchical group-lasso
  regularization}.{\BBCQ}
\newblock
\APACjournalVolNumPages{Journal of Computational and Graphical
  Statistics}{24}{3}{627--654}.
\PrintBackRefs{\CurrentBib}

\bibitem [\protect \citeauthoryear {%
Malyutov%
}{%
Malyutov%
}{%
{\protect \APACyear {2005}}%
}]{%
malyutov2005authorship}
\APACinsertmetastar {%
malyutov2005authorship}%
\begin{APACrefauthors}%
Malyutov, M\BPBI B.%
\end{APACrefauthors}%
\unskip\
\newblock
\APACrefYearMonthDay{2005}{}{}.
\newblock
{\BBOQ}\APACrefatitle {Authorship attribution of texts: a review} {Authorship
  attribution of texts: a review}.{\BBCQ}
\newblock
\APACjournalVolNumPages{Electronic Notes in Discrete
  Mathematics}{21}{}{353--357}.
\PrintBackRefs{\CurrentBib}

\bibitem [\protect \citeauthoryear {%
Manning%
\ \BBA {} Sch{\"u}tze%
}{%
Manning%
\ \BBA {} Sch{\"u}tze%
}{%
{\protect \APACyear {1999}}%
}]{%
manning1999foundations}
\APACinsertmetastar {%
manning1999foundations}%
\begin{APACrefauthors}%
Manning, C\BPBI D.%
\BCBT {}\ \BBA {} Sch{\"u}tze, H.%
\end{APACrefauthors}%
\unskip\
\newblock
\APACrefYear{1999}.
\newblock
\APACrefbtitle {Foundations of Statistical Natural Language Processing}
  {Foundations of statistical natural language processing}.
\newblock
\APACaddressPublisher{}{MIT Press}.
\PrintBackRefs{\CurrentBib}

\bibitem [\protect \citeauthoryear {%
McCormick%
}{%
McCormick%
}{%
{\protect \APACyear {1998}}%
}]{%
mccormick1998}
\APACinsertmetastar {%
mccormick1998}%
\begin{APACrefauthors}%
McCormick, N.%
\end{APACrefauthors}%
\unskip\
\newblock
\APACrefYearMonthDay{1998}{January}{10}.
\newblock
\APACrefbtitle {Must it be {L}ennon or {M}c{C}artney?} {Must it be {L}ennon or
  {M}c{C}artney?}
\newblock
\begin{APACrefURL}
  \url{http://www.telegraph.co.uk/culture/4711552/Must-it-be-Lennon-or-McCartney.html}
  \end{APACrefURL}
\newblock
\APACrefnote{Accessed 07-June-2017}
\PrintBackRefs{\CurrentBib}

\bibitem [\protect \citeauthoryear {%
McDougal%
}{%
McDougal%
}{%
{\protect \APACyear {2013}}%
}]{%
mcdougal2013}
\APACinsertmetastar {%
mcdougal2013}%
\begin{APACrefauthors}%
McDougal, C.%
\end{APACrefauthors}%
\unskip\
\newblock
\APACrefYearMonthDay{2013}{August}{}.
\newblock
\APACrefbtitle {Multi-dimensional computer-driven quantitative analysis of the
  music and lyrics of the {B}eatles} {Multi-dimensional computer-driven
  quantitative analysis of the music and lyrics of the {B}eatles}\
  \APACbVolEdTR {}{Technical report}.
\newblock
\APACaddressInstitution{}{Northeastern University}.
\newblock
\begin{APACrefURL} \url{https://cedricmcdougal.com/4/papers/beatles.pdf}
  \end{APACrefURL}
\PrintBackRefs{\CurrentBib}

\bibitem [\protect \citeauthoryear {%
Middleton%
}{%
Middleton%
}{%
{\protect \APACyear {1990}}%
}]{%
middleton1990studying}
\APACinsertmetastar {%
middleton1990studying}%
\begin{APACrefauthors}%
Middleton, R.%
\end{APACrefauthors}%
\unskip\
\newblock
\APACrefYear{1990}.
\newblock
\APACrefbtitle {Studying {P}opular {M}usic} {Studying {P}opular {M}usic}.
\newblock
\APACaddressPublisher{}{McGraw-Hill Education (UK)}.
\PrintBackRefs{\CurrentBib}

\bibitem [\protect \citeauthoryear {%
Miles%
}{%
Miles%
}{%
{\protect \APACyear {1998}}%
}]{%
miles1998paul}
\APACinsertmetastar {%
miles1998paul}%
\begin{APACrefauthors}%
Miles, B.%
\end{APACrefauthors}%
\unskip\
\newblock
\APACrefYear{1998}.
\newblock
\APACrefbtitle {Paul {M}c{C}artney: Many {Y}ears from {N}ow} {Paul
  {M}c{C}artney: Many {Y}ears from {N}ow}.
\newblock
\APACaddressPublisher{}{Macmillan}.
\PrintBackRefs{\CurrentBib}

\bibitem [\protect \citeauthoryear {%
Mosteller%
\ \BBA {} Wallace%
}{%
Mosteller%
\ \BBA {} Wallace%
}{%
{\protect \APACyear {1963}}%
}]{%
mosteller1963inference}
\APACinsertmetastar {%
mosteller1963inference}%
\begin{APACrefauthors}%
Mosteller, F.%
\BCBT {}\ \BBA {} Wallace, D\BPBI L.%
\end{APACrefauthors}%
\unskip\
\newblock
\APACrefYearMonthDay{1963}{}{}.
\newblock
{\BBOQ}\APACrefatitle {Inference in an authorship problem: A comparative study
  of discrimination methods applied to the authorship of the disputed
  Federalist Papers} {Inference in an authorship problem: A comparative study
  of discrimination methods applied to the authorship of the disputed
  federalist papers}.{\BBCQ}
\newblock
\APACjournalVolNumPages{Journal of the American Statistical
  Association}{58}{302}{275--309}.
\PrintBackRefs{\CurrentBib}

\bibitem [\protect \citeauthoryear {%
Mosteller%
\ \BBA {} Wallace%
}{%
Mosteller%
\ \BBA {} Wallace%
}{%
{\protect \APACyear {1984}}%
}]{%
mosteller1984applied}
\APACinsertmetastar {%
mosteller1984applied}%
\begin{APACrefauthors}%
Mosteller, F.%
\BCBT {}\ \BBA {} Wallace, D\BPBI L.%
\end{APACrefauthors}%
\unskip\
\newblock
\APACrefYear{1984}.
\newblock
\APACrefbtitle {Applied {B}ayesian and {C}lassical {I}nference: {T}he {C}ase of
  the {F}ederalist {P}apers} {Applied {B}ayesian and {C}lassical {I}nference:
  {T}he {C}ase of the {F}ederalist {P}apers}.
\newblock
\APACaddressPublisher{}{Springer}.
\PrintBackRefs{\CurrentBib}

\bibitem [\protect \citeauthoryear {%
Naccache%
, Borgi%
\BCBL {}\ \BBA {} Gh{\'e}dira%
}{%
Naccache%
\ \protect \BOthers {.}}{%
{\protect \APACyear {2008}}%
}]{%
naccache2008learning}
\APACinsertmetastar {%
naccache2008learning}%
\begin{APACrefauthors}%
Naccache, M.%
, Borgi, A.%
\BCBL {}\ \BBA {} Gh{\'e}dira, K.%
\end{APACrefauthors}%
\unskip\
\newblock
\APACrefYearMonthDay{2008}{}{}.
\newblock
{\BBOQ}\APACrefatitle {A Learning-Based Model for Musical Data Representation
  Using Histograms} {A learning-based model for musical data representation
  using histograms}.{\BBCQ}
\newblock
\BIn{} \APACrefbtitle {International Symposium on Computer Music Modeling and
  Retrieval} {International symposium on computer music modeling and
  retrieval}\ (\BPGS\ 207--215).
\PrintBackRefs{\CurrentBib}

\bibitem [\protect \citeauthoryear {%
Robin%
\ \protect \BOthers {.}}{%
Robin%
\ \protect \BOthers {.}}{%
{\protect \APACyear {2011}}%
}]{%
pROC2011}
\APACinsertmetastar {%
pROC2011}%
\begin{APACrefauthors}%
Robin, X.%
, Turck, N.%
, Hainard, A.%
, Tiberti, N.%
, Lisacek, F.%
, Sanchez, J\BHBI C.%
\BCBL {}\ \BBA {} M\"{u}ller, M.%
\end{APACrefauthors}%
\unskip\
\newblock
\APACrefYearMonthDay{2011}{}{}.
\newblock
{\BBOQ}\APACrefatitle {{p}{ROC}: An open-source package for {R} and {S}+ to
  analyze and compare {ROC} curves} {{p}{ROC}: An open-source package for {R}
  and {S}+ to analyze and compare {ROC} curves}.{\BBCQ}
\newblock
\APACjournalVolNumPages{BMC Bioinformatics}{12}{}{77}.
\PrintBackRefs{\CurrentBib}

\bibitem [\protect \citeauthoryear {%
{Rolling Stone}%
}{%
{Rolling Stone}%
}{%
{\protect \APACyear {2011}}%
}]{%
rolling2011}
\APACinsertmetastar {%
rolling2011}%
\begin{APACrefauthors}%
{Rolling Stone}.%
\end{APACrefauthors}%
\unskip\
\newblock
\APACrefYearMonthDay{2011}{April}{}.
\newblock
\APACrefbtitle {The {B}eatles, {In My Life}.} {The {B}eatles, {In My Life}.}
\newblock
\begin{APACrefURL}
  \url{https://www.rollingstone.com/music/music-lists/500-greatest-songs-of-all-time-151127/the-beatles-in-my-life-57758/}
  \end{APACrefURL}
\newblock
\APACrefnote{Accessed 19-August-2018}
\PrintBackRefs{\CurrentBib}

\bibitem [\protect \citeauthoryear {%
Ruczinski%
, Kooperberg%
\BCBL {}\ \BBA {} LeBlanc%
}{%
Ruczinski%
\ \protect \BOthers {.}}{%
{\protect \APACyear {2003}}%
}]{%
ruczinski2003logic}
\APACinsertmetastar {%
ruczinski2003logic}%
\begin{APACrefauthors}%
Ruczinski, I.%
, Kooperberg, C.%
\BCBL {}\ \BBA {} LeBlanc, M.%
\end{APACrefauthors}%
\unskip\
\newblock
\APACrefYearMonthDay{2003}{}{}.
\newblock
{\BBOQ}\APACrefatitle {Logic regression} {Logic regression}.{\BBCQ}
\newblock
\APACjournalVolNumPages{Journal of Computational and Graphical
  Statistics}{12}{3}{475--511}.
\PrintBackRefs{\CurrentBib}

\bibitem [\protect \citeauthoryear {%
Ruczinski%
, Kooperberg%
\BCBL {}\ \BBA {} LeBlanc%
}{%
Ruczinski%
\ \protect \BOthers {.}}{%
{\protect \APACyear {2004}}%
}]{%
ruczinski2004exploring}
\APACinsertmetastar {%
ruczinski2004exploring}%
\begin{APACrefauthors}%
Ruczinski, I.%
, Kooperberg, C.%
\BCBL {}\ \BBA {} LeBlanc, M\BPBI L.%
\end{APACrefauthors}%
\unskip\
\newblock
\APACrefYearMonthDay{2004}{}{}.
\newblock
{\BBOQ}\APACrefatitle {Exploring interactions in high-dimensional genomic data:
  an overview of logic regression, with applications} {Exploring interactions
  in high-dimensional genomic data: an overview of logic regression, with
  applications}.{\BBCQ}
\newblock
\APACjournalVolNumPages{Journal of Multivariate Analysis}{90}{1}{178--195}.
\PrintBackRefs{\CurrentBib}

\bibitem [\protect \citeauthoryear {%
Rybaczewski%
}{%
Rybaczewski%
}{%
{\protect \APACyear {2018}}%
}]{%
rybac2018}
\APACinsertmetastar {%
rybac2018}%
\begin{APACrefauthors}%
Rybaczewski, D.%
\end{APACrefauthors}%
\unskip\
\newblock
\APACrefYearMonthDay{2018}{}{}.
\newblock
\APACrefbtitle {A {H}ard {D}ay's {N}ight.} {A {H}ard {D}ay's {N}ight.}
\newblock
\begin{APACrefURL} \url{http://www.beatlesebooks.com/hard-days-night}
  \end{APACrefURL}
\newblock
\APACrefnote{Accessed 19-August-2018}
\PrintBackRefs{\CurrentBib}

\bibitem [\protect \citeauthoryear {%
Thisted%
\ \BBA {} Efron%
}{%
Thisted%
\ \BBA {} Efron%
}{%
{\protect \APACyear {1987}}%
}]{%
thisted1987did}
\APACinsertmetastar {%
thisted1987did}%
\begin{APACrefauthors}%
Thisted, R.%
\BCBT {}\ \BBA {} Efron, B.%
\end{APACrefauthors}%
\unskip\
\newblock
\APACrefYearMonthDay{1987}{}{}.
\newblock
{\BBOQ}\APACrefatitle {Did {S}hakespeare write a newly-discovered poem?} {Did
  {S}hakespeare write a newly-discovered poem?}{\BBCQ}
\newblock
\APACjournalVolNumPages{Biometrika}{74}{3}{445--455}.
\PrintBackRefs{\CurrentBib}

\bibitem [\protect \citeauthoryear {%
Tibshirani%
}{%
Tibshirani%
}{%
{\protect \APACyear {1996}}%
}]{%
tibshirani1996regression}
\APACinsertmetastar {%
tibshirani1996regression}%
\begin{APACrefauthors}%
Tibshirani, R.%
\end{APACrefauthors}%
\unskip\
\newblock
\APACrefYearMonthDay{1996}{}{}.
\newblock
{\BBOQ}\APACrefatitle {Regression shrinkage and selection via the lasso}
  {Regression shrinkage and selection via the lasso}.{\BBCQ}
\newblock
\APACjournalVolNumPages{Journal of the Royal Statistical Society, Series B
  (Statistical Methodology)}{58}{1}{267--288}.
\PrintBackRefs{\CurrentBib}

\bibitem [\protect \citeauthoryear {%
Tibshirani%
}{%
Tibshirani%
}{%
{\protect \APACyear {2011}}%
}]{%
tibshirani2011regression}
\APACinsertmetastar {%
tibshirani2011regression}%
\begin{APACrefauthors}%
Tibshirani, R.%
\end{APACrefauthors}%
\unskip\
\newblock
\APACrefYearMonthDay{2011}{}{}.
\newblock
{\BBOQ}\APACrefatitle {Regression shrinkage and selection via the lasso: a
  retrospective} {Regression shrinkage and selection via the lasso: a
  retrospective}.{\BBCQ}
\newblock
\APACjournalVolNumPages{Journal of the Royal Statistical Society, Series B
  (Statistical Methodology)}{73}{3}{273--282}.
\PrintBackRefs{\CurrentBib}

\bibitem [\protect \citeauthoryear {%
Turner%
}{%
Turner%
}{%
{\protect \APACyear {1999}}%
}]{%
turner1999hard}
\APACinsertmetastar {%
turner1999hard}%
\begin{APACrefauthors}%
Turner, S.%
\end{APACrefauthors}%
\unskip\
\newblock
\APACrefYear{1999}.
\newblock
\APACrefbtitle {A {H}ard {D}ay's {W}rite: {T}he stories behind every {B}eatles
  song} {A {H}ard {D}ay's {W}rite: {T}he stories behind every {B}eatles song}.
\newblock
\APACaddressPublisher{}{Carlton, Dubai}.
\PrintBackRefs{\CurrentBib}

\bibitem [\protect \citeauthoryear {%
Wagner%
}{%
Wagner%
}{%
{\protect \APACyear {2003}}%
}]{%
wagner2003domestication}
\APACinsertmetastar {%
wagner2003domestication}%
\begin{APACrefauthors}%
Wagner, N.%
\end{APACrefauthors}%
\unskip\
\newblock
\APACrefYearMonthDay{2003}{}{}.
\newblock
{\BBOQ}\APACrefatitle {``{D}omestication'' of Blue Notes in the {B}eatles'
  Songs} {``{D}omestication'' of blue notes in the {B}eatles' songs}.{\BBCQ}
\newblock
\APACjournalVolNumPages{Music Theory Spectrum}{25}{2}{353--365}.
\PrintBackRefs{\CurrentBib}

\bibitem [\protect \citeauthoryear {%
Wenner%
}{%
Wenner%
}{%
{\protect \APACyear {2009}}%
}]{%
wenner2009}
\APACinsertmetastar {%
wenner2009}%
\begin{APACrefauthors}%
Wenner, J.%
\end{APACrefauthors}%
\unskip\
\newblock
\APACrefYearMonthDay{2009}{}{}.
\newblock
\APACrefbtitle {John {L}ennon {R}emembers - {J}ann {W}enner {I}nterview {P}art
  5.} {John {L}ennon {R}emembers - {J}ann {W}enner {I}nterview {P}art 5.}
\newblock
\begin{APACrefURL}
  \url{http://tittenhurstlennon.blogspot.com/2009/07/jann-wenner-interview-part-5.html}
  \end{APACrefURL}
\newblock
\APACrefnote{Accessed 14-January-2019}
\PrintBackRefs{\CurrentBib}

\bibitem [\protect \citeauthoryear {%
Whissell%
}{%
Whissell%
}{%
{\protect \APACyear {1996}}%
}]{%
whissell1996traditional}
\APACinsertmetastar {%
whissell1996traditional}%
\begin{APACrefauthors}%
Whissell, C.%
\end{APACrefauthors}%
\unskip\
\newblock
\APACrefYearMonthDay{1996}{}{}.
\newblock
{\BBOQ}\APACrefatitle {Traditional and emotional stylometric analysis of the
  songs of {B}eatles {P}aul {M}c{C}artney and {J}ohn {L}ennon} {Traditional and
  emotional stylometric analysis of the songs of {B}eatles {P}aul {M}c{C}artney
  and {J}ohn {L}ennon}.{\BBCQ}
\newblock
\APACjournalVolNumPages{Computers and the Humanities}{30}{3}{257--265}.
\PrintBackRefs{\CurrentBib}

\bibitem [\protect \citeauthoryear {%
Wiener%
}{%
Wiener%
}{%
{\protect \APACyear {1986}}%
}]{%
wiener1986beatles}
\APACinsertmetastar {%
wiener1986beatles}%
\begin{APACrefauthors}%
Wiener, A\BPBI J.%
\end{APACrefauthors}%
\unskip\
\newblock
\APACrefYear{1986}.
\newblock
\APACrefbtitle {The {B}eatles: {A Recording History}} {The {B}eatles: {A
  Recording History}}.
\newblock
\APACaddressPublisher{}{McFarland \& Co Inc Pub}.
\PrintBackRefs{\CurrentBib}

\bibitem [\protect \citeauthoryear {%
Womack%
}{%
Womack%
}{%
{\protect \APACyear {2007}}%
}]{%
womack2007authorship}
\APACinsertmetastar {%
womack2007authorship}%
\begin{APACrefauthors}%
Womack, K.%
\end{APACrefauthors}%
\unskip\
\newblock
\APACrefYearMonthDay{2007}{}{}.
\newblock
{\BBOQ}\APACrefatitle {Authorship and the {B}eatles} {Authorship and the
  {B}eatles}.{\BBCQ}
\newblock
\APACjournalVolNumPages{College Literature}{34}{3}{161--182}.
\PrintBackRefs{\CurrentBib}

\bibitem [\protect \citeauthoryear {%
Yuan%
\ \BBA {} Lin%
}{%
Yuan%
\ \BBA {} Lin%
}{%
{\protect \APACyear {2006}}%
}]{%
yuan2006model}
\APACinsertmetastar {%
yuan2006model}%
\begin{APACrefauthors}%
Yuan, M.%
\BCBT {}\ \BBA {} Lin, Y.%
\end{APACrefauthors}%
\unskip\
\newblock
\APACrefYearMonthDay{2006}{}{}.
\newblock
{\BBOQ}\APACrefatitle {Model selection and estimation in regression with
  grouped variables} {Model selection and estimation in regression with grouped
  variables}.{\BBCQ}
\newblock
\APACjournalVolNumPages{Journal of the Royal Statistical Society, Series B
  (Statistical Methodology)}{68}{1}{49--67}.
\PrintBackRefs{\CurrentBib}

\bibitem [\protect \citeauthoryear {%
Zou%
\ \BBA {} Hastie%
}{%
Zou%
\ \BBA {} Hastie%
}{%
{\protect \APACyear {2005}}%
}]{%
zou2005regularization}
\APACinsertmetastar {%
zou2005regularization}%
\begin{APACrefauthors}%
Zou, H.%
\BCBT {}\ \BBA {} Hastie, T.%
\end{APACrefauthors}%
\unskip\
\newblock
\APACrefYearMonthDay{2005}{}{}.
\newblock
{\BBOQ}\APACrefatitle {Regularization and variable selection via the elastic
  net} {Regularization and variable selection via the elastic net}.{\BBCQ}
\newblock
\APACjournalVolNumPages{Journal of the Royal Statistical Society, Series B
  (Statistical Methodology)}{67}{2}{301--320}.
\PrintBackRefs{\CurrentBib}

\end{thebibliography}

\end{document}